\documentstyle[aps,epsf,citesort,rotate,multicol]{revtex}

\begin{document}

\draft

\title{A Random Matrix Approach to Cross-Correlations in Financial~Data}

\author{Vasiliki Plerou$^{1,2}$\footnote{Email plerou@cgl.bu.edu (corresponding author)}, Parameswaran Gopikrishnan$^{1}$, Bernd Rosenow$^{1,3}$,\\ Lu\'{\i}s~ A. Nunes Amaral$^{1}$, Thomas Guhr$^{4}$, and H. Eugene Stanley$^{1}$}

\address{ $^{1}$ Center for Polymer Studies and Department of Physics, Boston
        University, Boston, Massachusetts 02215, USA \\
        $^{2}$Department of Physics, Boston College, Chestnut Hill,
        Massachusetts 02167, USA \\ $^{3}$ Department of Physics,
        Harvard University, Cambridge, Massachusetts 02138, USA \\
        $^{4}$ Max--Planck--Institute for Nuclear Physics, D--69029
        Heidelberg, Germany\\ }

\date{July 31, 2001.}

\maketitle

\begin{abstract}

We analyze cross-correlations between price fluctuations of different
stocks using methods of random matrix theory (RMT). Using two large
databases, we calculate cross-correlation matrices {\bf \sf C} of
returns constructed from (i) 30-min returns of 1000 US stocks for the
2-yr period 1994--95 (ii) 30-min returns of 881 US stocks for the 2-yr
period 1996--97, and (iii) 1-day returns of 422 US stocks for the
35-yr period 1962--96. We test the statistics of the eigenvalues
$\lambda_i$ of {\bf \sf C} against a ``null hypothesis'' --- a random
correlation matrix constructed from mutually uncorrelated time
series. We find that a majority of the eigenvalues of {\bf \sf C} fall
within the RMT bounds $[\lambda_- , \lambda_+]$ for the eigenvalues of
random correlation matrices. We test the eigenvalues of {\bf \sf C}
within the RMT bound for universal properties of random matrices and
find good agreement with the results for the Gaussian orthogonal
ensemble of random matrices --- implying a large degree of randomness
in the measured cross-correlation coefficients. Further, we find that
the distribution of eigenvector components for the eigenvectors
corresponding to the eigenvalues outside the RMT bound display
systematic deviations from the RMT prediction. In addition, we find
that these ``deviating eigenvectors'' are stable in time. We analyze
the components of the deviating eigenvectors and find that the largest
eigenvalue corresponds to an influence common to all stocks. Our
analysis of the remaining deviating eigenvectors shows distinct
groups, whose identities correspond to conventionally-identified
business sectors. Finally, we discuss applications to the construction
of portfolios of stocks that have a stable ratio of risk to return.

\end{abstract}
\pacs{PACS numbers: 05.45.Tp, 89.90.+n, 05.40.-a, 05.40.Fb}
\begin{multicols}{2}

\section{Introduction}

\subsection{Motivation}

Quantifying correlations between different stocks is a topic of
interest not only for scientific reasons of understanding the economy
as a complex dynamical system, but also for practical reasons such as
asset allocation and portfolio-risk
estimation~\cite{Mantegna99,Bouchaud00,Farmer99,CLM}. Unlike most
physical systems, where one relates correlations between subunits to
basic interactions, the underlying ``interactions'' for the stock
market problem are not known. Here, we analyze cross-correlations
between stocks by applying concepts and methods of random matrix
theory, developed in the context of complex quantum systems where the
precise nature of the interactions between subunits are not known.

In order to quantify correlations, we first calculate the price change
(``return'') of stock $i=1,\dots,N$ over a time scale $\Delta t$
\begin{equation}
G_i(t)\equiv \ln S_i(t+\Delta t) - \ln S_i(t)\,,
\label{return}
\end{equation}
where $S_i(t)$ denotes the price of stock $i$. Since different stocks
have varying levels of volatility (standard deviation), we define a
normalized return
\begin{equation}
g_i(t)\equiv {G_i(t) - \langle G_i \rangle \over \sigma_i}\,,
\label{norm-ret}
\end{equation}
where $\sigma_i \equiv \sqrt{\langle G_i^2 \rangle - \langle G_i
\rangle^2} $ is the standard deviation of $G_i$, and
$\langle\cdots\rangle$ denotes a time average over the period
studied. We then compute the equal-time cross-correlation matrix {\bf
\sf C} with elements
%
\begin{equation}
C_{ij} \equiv \langle g_i(t) g_j(t) \rangle\,.
\label{eq.2}
\end{equation}
%
By construction, the elements $C_{ij}$ are restricted to the domain
$-1 \leq C_{ij} \leq 1$, where $C_{ij}=1$ corresponds to perfect
correlations, $C_{ij}=-1$ corresponds to perfect anti-correlations, and
$C_{ij}=0$ corresponds to uncorrelated pairs of stocks.

The difficulties in analyzing the significance and meaning of the
empirical cross-correlation coefficients $C_{ij}$ are due to several
reasons, which include the following: 

\vspace{0.5cm}

\noindent (i) Market conditions change with time and the cross-correlations 
that exist between any pair of stocks may not be stationary.

\vspace{0.5cm}

\noindent (ii) The finite length of time series available to estimate
cross-correlations introduces ``measurement noise''.

\vspace{0.5cm}

\noindent If we use a long time series to circumvent the problem of finite
length, our estimates will be affected by the non-stationarity of
cross-correlations. For these reasons, the empirically-measured
cross-correlations will contain ``random'' contributions, and it is a
difficult problem in general to estimate from {\bf \sf C} the
cross-correlations that are not a result of randomness.

How can we identify from $C_{ij}$, those stocks that remained
correlated (on the average) in the time period studied?  To answer
this question, we test the statistics of {\bf \sf C} against the
``null hypothesis'' of a random correlation matrix --- a correlation
matrix constructed from mutually uncorrelated time series. If the
properties of {\bf \sf C} conform to those of a random correlation
matrix, then it follows that the contents of the empirically-measured
{\bf \sf C} are random. Conversely, deviations of the properties of
{\bf \sf C} from those of a random correlation matrix convey
information about ``genuine'' correlations. Thus, our goal shall be to
compare the properties of {\bf \sf C} with those of a random
correlation matrix and separate the content of {\bf \sf C} into two
groups: (a) the part of {\bf \sf C} that conforms to the properties of
random correlation matrices (``noise'') and (b) the part of {\bf
\sf C} that deviates (``information'').

\subsection{Background}

The study of statistical properties of matrices with independent
random elements --- {\it random matrices} --- has a rich history
originating in nuclear
physics~\cite{wigner,wigner2,wigner3,dyson,Dyson63,Mehta63,Mehta91,Brody81,Guhr98}. In
nuclear physics, the problem of interest 50~years ago was to
understand the energy levels of complex nuclei, which the existing
models failed to explain. RMT was developed in this context by Wigner,
Dyson, Mehta, and others in order to explain the statistics of energy
levels of complex quantum systems. They postulated that the
Hamiltonian describing a heavy nucleus can be described by a matrix
{\bf \sf H} with independent random elements $H_{ij}$ drawn from a
probability distribution~\cite{wigner,wigner2,wigner3,dyson,Dyson63}.
Based on this assumption, a series of remarkable predictions were made
which are found to be in agreement with the experimental
data~\cite{wigner,wigner2,wigner3}.  For complex quantum systems, RMT
predictions represent an average over all possible
interactions~\cite{dyson,Dyson63,Mehta63}. Deviations from the {\it
universal} predictions of RMT identify system-specific, non-random
properties of the system under consideration, providing clues about
the underlying interactions \cite{Mehta91,Brody81,Guhr98}.

Recent studies~\cite{Laloux99,Plerou99} applying RMT methods to
analyze the properties of {\bf \sf C} show that $\approx 98$\% of the
eigenvalues of {\bf \sf C} agree with RMT predictions, suggesting a
considerable degree of randomness in the measured
cross-correlations. It is also found that there are deviations from
RMT predictions for $\approx 2$\% of the largest eigenvalues. These
results prompt the following questions:
\begin{itemize}

\item
What is a possible interpretation for the deviations from RMT?

\item
Are the deviations from RMT stable in time? 

\item
What can we infer about the structure of {\bf \sf C} from these results? 

\item
What are the practical implications of these results?

\end{itemize}

In the following, we address these questions in detail. We find that
the largest eigenvalue of {\bf \sf C} represents the influence of the
entire market that is common to all stocks. Our analysis of the
contents of the remaining eigenvalues that deviate from RMT shows the
existence of cross-correlations between stocks of the same type of
industry, stocks having large market capitalization, and stocks of
firms having business in certain geographical
areas~\cite{ultra,Gopi00}. By calculating the scalar product of the
eigenvectors from one time period to the next, we find that the
``deviating eigenvectors'' have varying degrees of time stability,
quantified by the magnitude of the scalar product. The largest 2-3
eigenvectors are stable for extended periods of time, while for the
rest of the deviating eigenvectors, the time stability decreases as
the the corresponding eigenvalues are closer to the RMT upper bound.

To test that the deviating eigenvalues are the only ``genuine''
information contained in {\bf \sf C}, we compare the eigenvalue
statistics of {\bf \sf C} with the known universal properties of real
symmetric random matrices, and we find good agreement with the RMT
results. Using the notion of the inverse participation ratio, we
analyze the eigenvectors of {\bf \sf C} and find large values of
inverse participation ratio at both edges of the eigenvalue spectrum
--- suggesting a ``random band'' matrix structure for {\bf \sf
C}. Lastly, we discuss applications to the practical goal of finding
an investment that provides a given return without exposure to
unnecessary risk. In addition, it is possible that our methods can
also be applied for filtering out `noise' in empirically-measured
cross-correlation matrices in a wide variety of applications.

This paper is organized as follows. Section II contains a brief
description of the data analyzed. Section III discusses the statistics
of cross-correlation coefficients. Section IV discusses the eigenvalue
distribution of {\bf \sf C} and compares with RMT results. Section V
tests the eigenvalue statistics {\bf \sf C} for universal properties
of real symmetric random matrices and Section VI contains a detailed
analysis of the contents of eigenvectors that deviate from
RMT. Section VII discusses the time stability of the deviating
eigenvectors. Section VIII contains applications of RMT methods to
construct `optimal' portfolios that have a stable ratio of risk to
return. Finally, Section IX contains some concluding remarks.

\section{Data Analyzed}

We analyze two different databases covering securities from the three
major US stock exchanges, namely the New York Stock Exchange (NYSE),
the American Stock Exchange (AMEX), and the National Association of
Securities Dealers Automated Quotation (Nasdaq). 

\vspace{0.5cm}

\noindent $\bullet$ {\bf Database I:} We analyze the Trades and 
Quotes database, that documents all transactions for all major
securities listed in all the three stock exchanges. We extract from
this database time series of prices~\cite{splits} of the 1000 largest
stocks by market capitalization on the starting date January 3,
1994. We analyze this database for the 2-yr period
1994--95~\cite{notedata}. From this database, we form $L=6448$ records
of 30-min returns of $N=1000$ US stocks for the 2-yr period
1994--95. We also analyze the prices of a subset comprising 881 stocks
(of those 1000 we analyze for 1994--95) that survived through two
additional years 1996--97. From this data, we extract $L=6448$ records
of 30-min returns of $N=881$ US stocks for the 2-yr period 1996--97.

\vspace{0.5cm}

\noindent $\bullet$ {\bf Database II:} We analyze the Center for 
Research in Security Prices (CRSP) database.  The CRSP stock files
cover common stocks listed on NYSE beginning in 1925, the AMEX
beginning in 1962, and the Nasdaq beginning in 1972. The files provide
complete historical descriptive information and market data including
comprehensive distribution information, high, low and closing prices,
trading volumes, shares outstanding, and total returns.  We analyze
daily returns for the stocks that survive for the 35-yr period
1962--96 and extract $L=8685$ records of 1-day returns for $N=422$
stocks.

\section{Statistics of correlation coefficients}

\begin{figure}[h]
\narrowtext 
\centerline{
\epsfysize=0.7\columnwidth{\rotate[r]{\epsfbox{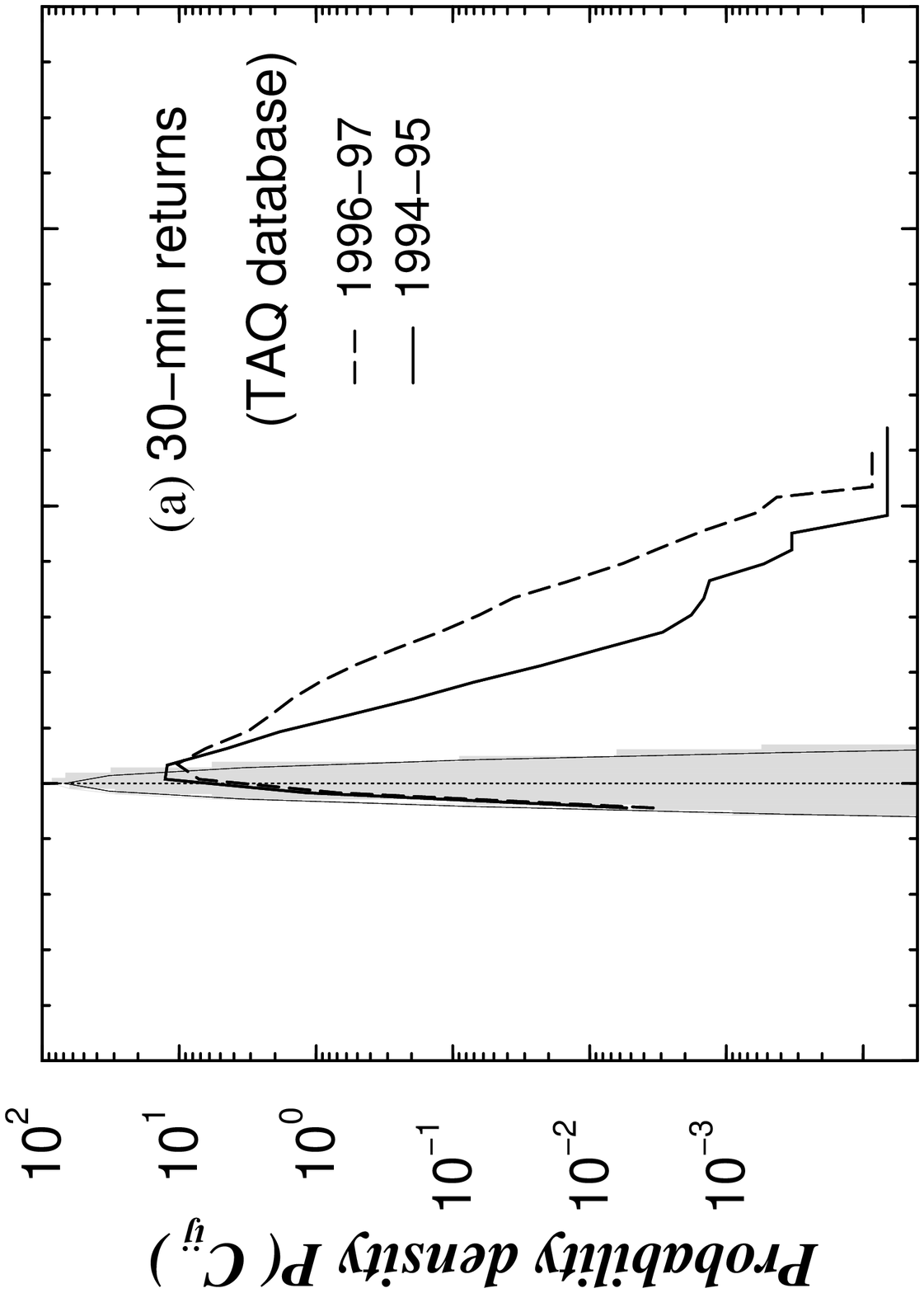}}} 
}
\centerline{
\epsfysize=0.7\columnwidth{\rotate[r]{\epsfbox{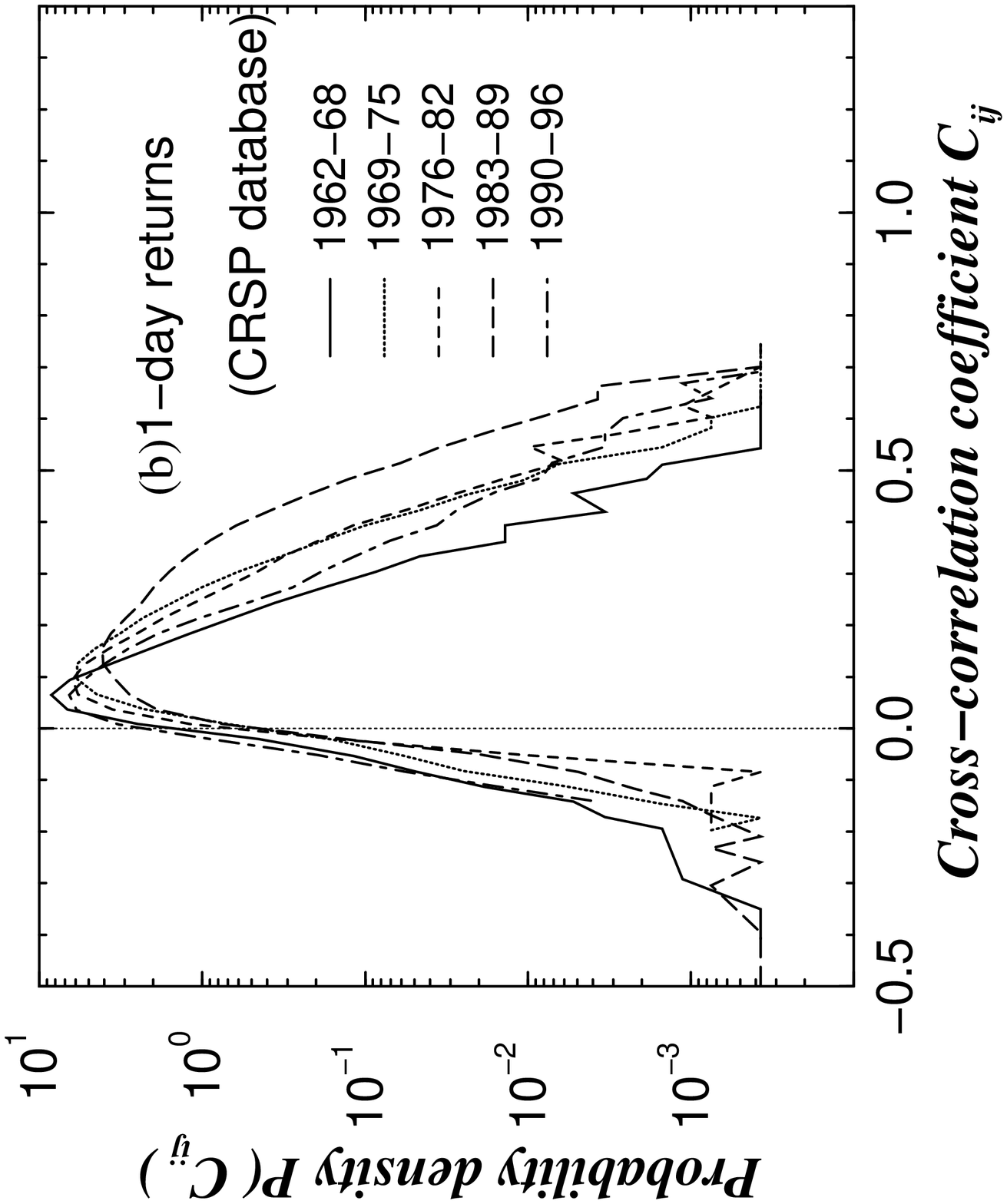}}} 
}
\caption{(a) $P(C_{ij})$ for {\bf \sf C} calculated using 
30-min returns of 1000 stocks for the 2-yr period 1994--95 (solid line)
and 881 stocks for the 2-yr period 1996--97 (dashed line). For the
period 1996--97 $\langle C_{ij} \rangle = 0.06$, larger than the value
$\langle C_{ij} \rangle = 0.03$ for 1994--95. The shaded region shows
the distribution of correlation coefficients for the control
$P(R_{ij})$ of Eq.~(\protect\ref{matrixA}), which is consistent with a
Gaussian distribution with zero mean. (b) $P(C_{ij})$ calculated from
daily returns of 422 stocks for five 7-yr sub-periods in the 35 years
1962--96. We find a large value of $\langle C_{ij} \rangle=0.18$ for
the period 1983--89, compared with the average $\langle C_{ij}
\rangle=0.10$ for the other periods.}
\label{distcij}
\end{figure}

We analyze the distribution $P(C_{ij})$ of the elements $\{C_{ij};\, i
\ne j\}$ of the cross-correlation matrix {\bf \sf C }. We first examine
$P(C_{ij})$ for 30-min returns from the TAQ database for the 2-yr
periods 1994--95 and 1996--97 [Fig.~\ref{distcij}(a)]. First, we note
that $P(C_{ij})$ is asymmetric and centered around a positive mean
value ($\langle C_{ij} \rangle > 0$), implying that
positively-correlated behavior is more prevalent than
negatively-correlated (anti-correlated) behavior. Secondly, we find
that $\langle C_{ij} \rangle$ depends on time, e.g., the period
1996--97 shows a larger $\langle C_{ij}
\rangle$ than the period 1994--95.  We contrast $P(C_{ij})$ with a
control --- a correlation matrix {\bf \sf R} with elements $R_{ij}$
constructed from $N=1000$ mutually-uncorrelated time series, each of
length $L=6448$, generated using the empirically-found distribution of
stock returns~\cite{Plerou99b,Lux96}. Figure~\ref{distcij}(a) shows
that $P(R_{ij})$ is consistent with a Gaussian with zero mean, in
contrast to $P(C_{ij})$. In addition, we see that the part of
$P(C_{ij})$ for $C_{ij} <0$ (which corresponds to anti-correlations)
is within the Gaussian curve for the control, suggesting the
possibility that the observed negative cross-correlations in {\bf \sf
C} may be an effect of randomness.

\begin{figure}[hbt]
\narrowtext 
\centerline{
\epsfysize=0.7\columnwidth{\rotate[r]{\epsfbox{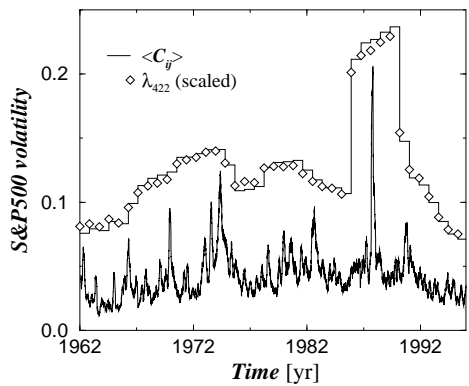}}} 
}
\caption{The stair-step curve shows the average value of the correlation 
coefficients $\langle C_{ij} \rangle$, calculated from $422 \times
422$ correlation matrices {\bf \sf C} constructed from daily returns
using a sliding $L=965$~day time window in discrete steps of
$L/5=193$~days. The diamonds correspond to the largest eigenvalue
$\lambda_{422}$ (scaled by a factor $4\times 10^2$) for the
correlation matrices thus obtained. The bottom curve shows the S\&P
500 volatility (scaled for clarity) calculated from daily records with
a sliding window of length 40~days. We find that both $\langle C_{ij}
\rangle$ and $\lambda_{422}$ have large values for periods containing
the market crash of October 19, 1987.}
\label{volat}
\end{figure}

Figure~\ref{distcij}(b) shows $P(C_{ij})$ for daily returns from the
CRSP database for five non-overlapping 7-yr sub-periods in the 35-yr
period 1962--96. We see that the time dependence of $\langle C_{ij}
\rangle$ is more pronounced in this plot. In particular, the period
containing the market crash of October 19, 1987 has the largest
average value $\langle C_{ij} \rangle$, suggesting the existence of
cross-correlations that are more pronounced in volatile periods than
in calm periods. We test this possibility by comparing $\langle C_{ij}
\rangle$ with the average volatility of the market (measured using the
S\&P 500 index), which shows large values of $\langle C_{ij} \rangle$
during periods of large volatility [Fig.~\ref{volat}].

\section{Eigenvalue distribution of the correlation matrix} 

As stated above, our aim is to extract information about
cross-correlations from {\bf \sf C}. So, we compare the properties of
{\bf \sf C} with those of a random cross-correlation
matrix~\cite{Laloux99}. In matrix notation, the correlation matrix can
be expressed as
\begin{equation}
{\bf \sf C = {\rm \it 1 \over L}\, G\, G^{T}}\,, 
\label{matrixC}
\end{equation}
where {\bf \sf G} is an $N\times L$ matrix with elements $\{g_{i\,m}
\equiv g_i(m\Delta t)\,;i=1,\dots,N\,; m=0,\dots,L-1\}\,$, and {\bf \sf
G$^T$} denotes the transpose of {\bf \sf G}. Therefore, we consider a
``random'' correlation matrix
\begin{equation}
{\bf \sf R = {\rm \it 1 \over L}\, A\, A^{T}}\,, 
\label{matrixA}
\end{equation}
where {\bf \sf A} is an $N\times L$ matrix containing $N$ time series
of $L$ random elements with zero mean and unit variance, that are
mutually uncorrelated. By construction {\bf \sf R} belongs to the
type of matrices often referred to as Wishart matrices in multivariate
statistics~\cite{Muirhead}.

Statistical properties of random matrices such as {\bf \sf R} are
known~\cite{Dyson71,Sengupta99}. Particularly, in the limit $N
\rightarrow \infty \,, L \rightarrow \infty$, such that $Q\equiv L/N$
is fixed, it was shown analytically~\cite{Sengupta99} that the
distribution $P_{\rm rm} (\lambda)$ of eigenvalues $\lambda$ of the
random correlation matrix {\bf \sf R} is given by
\begin{equation}
P_{\rm rm}(\lambda) = {Q \over 2 \pi}\, { \sqrt{(\lambda_{+} - \lambda)(\lambda - \lambda_{-})} \over \lambda}\, \,\,,
\label{densuncorr}
\end{equation}
for $\lambda$ within the bounds $\lambda_{-}\leq \lambda_i \leq
\lambda_{+}$, where $\lambda_-$ and $\lambda_+$ are the minimum and maximum 
eigenvalues of {\bf \sf R} respectively, given by
\begin{equation}
\lambda_{\pm} = 1 + {1 \over Q} \pm 2\, \sqrt{{1 \over Q}}\,.
\label{lambdamaxmin}
\end{equation}
For finite $L$ and $N$, the abrupt cut-off of $P_{\rm rm} (\lambda)$ is
replaced by a rapidly-decaying edge~\cite{Bowick91}.

\begin{figure}[h]
\narrowtext 
\centerline{
\epsfysize=0.9\columnwidth{\rotate[r]{\epsfbox{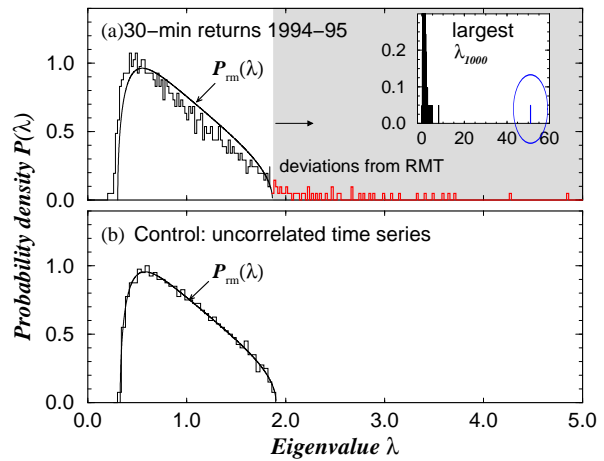}}} 
}
\vspace{0.5cm}
\caption{(a) Eigenvalue distribution $P(\lambda)$ for {\bf \sf C} 
constructed from the 30-min returns for 1000 stocks for the 2-yr
period 1994--95. The solid curve shows the RMT result $P_{\rm
rm}(\lambda)$ of Eq.~(\protect\ref{densuncorr}). We note several
eigenvalues outside the RMT upper bound $\lambda_+$ (shaded region).
The inset shows the largest eigenvalue $\lambda_{1000} \approx 50 \gg
\lambda_+$.  (b) $P(\lambda)$ for the random correlation matrix 
{\bf \sf R}, computed from $N=1000$ computer-generated random
uncorrelated time series with length $L=6448$ shows good agreement
with the RMT result, Eq.~(\protect\ref{densuncorr}) (solid curve).}
\label{evdist}
\end{figure}

We next compare the eigenvalue distribution $P(\lambda)$ of {\bf \sf
C} with $P_{\rm rm}(\lambda)$~\cite{Laloux99}. We examine $\Delta t =
30$~min returns for $N=1000$ stocks, each containing $L=6448$ records.
Thus $Q=6.448$, and we obtain $\lambda_-=0.36$ and $\lambda_+=1.94$
from Eq.~(\ref{lambdamaxmin}). We compute the eigenvalues $\lambda_i$
of {\bf \sf C}, where $\lambda_i$ are rank ordered ($\lambda_{i+1} >
\lambda_i$). Figure~\ref{evdist}(a) compares the probability 
distribution $P(\lambda)$ with $P_{\rm rm}(\lambda)$ calculated for
$Q=6.448$. We note the presence of a well-defined ``bulk'' of
eigenvalues which fall within the bounds $[\lambda_-,\lambda_+]$ for
$P_{\rm rm}(\lambda)$. We also note deviations for a few ($\approx
20$) largest and smallest eigenvalues. In particular, the largest
eigenvalue $\lambda_{1000} \approx 50$ for the 2-yr period, which is
$\approx 25$ times larger than $\lambda_{+}=1.94$.

Since Eq.~(\ref{densuncorr}) is strictly valid only for $L\rightarrow
\infty$ and $N\rightarrow \infty$, we must test that the deviations
that we find in Fig.~\ref{evdist}(a) for the largest few eigenvalues
are not an effect of finite values of $L$ and $N$.  To this end, we
contrast $P(\lambda)$ with the RMT result $P_{\rm rm} (\lambda)$ for
the random correlation matrix of Eq.~(\ref{matrixA}), constructed from
$N=1000$ separate uncorrelated time series, each of the same length
$L=6448$. We find good agreement with Eq.~(\ref{densuncorr})
[Fig.~\ref{evdist}(b)], thus showing that the deviations from RMT
found for the largest few eigenvalues in Fig.~\ref{evdist}(a) are not
a result of the fact that $L$ and $N$ are finite.

\begin{figure}[h]
\narrowtext 
\centerline{
\epsfysize=0.7\columnwidth{\rotate[r]{\epsfbox{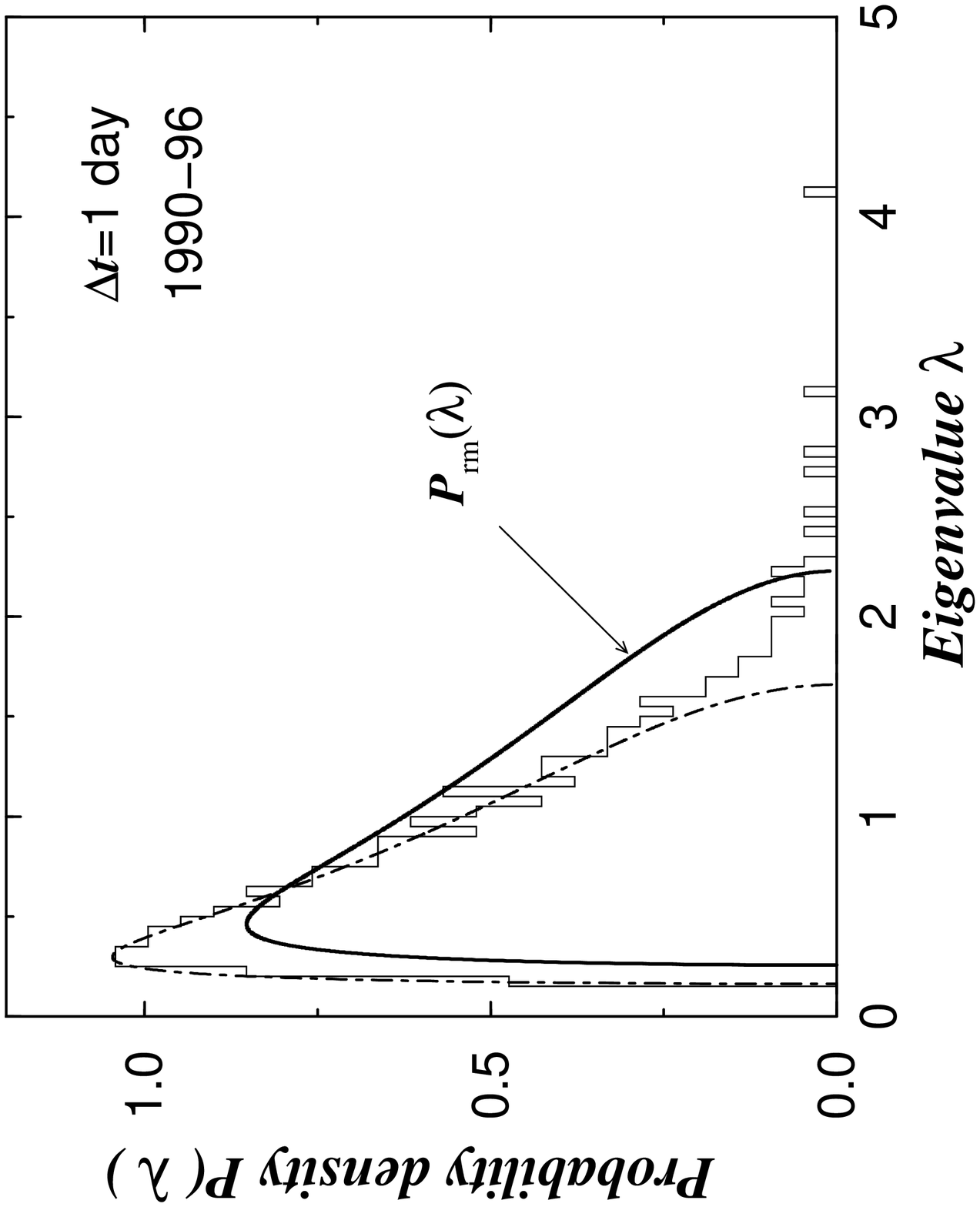}}} 
}
\vspace{0.5cm}
\caption{$P(\lambda)$ for {\bf  \sf C} constructed from daily returns of 
422 stocks for the 7-yr period 1990--96. The solid curve shows the RMT
result $P_{\rm rm} (\lambda)$ of Eq.~(\protect\ref{densuncorr}) using
$N=422$ and $L=1,737$. The dot-dashed curve shows a fit to
$P(\lambda)$ using $P_{\rm rm}(\lambda)$ with $\lambda_+$ and
$\lambda_-$ as free parameters. We find similar results as found in
Fig.~\ref{evdist}(a) for 30-min returns. The largest eigenvalue (not
shown) has the value $\lambda_{422}=46.3$.}
\label{evdist-daily}
\end{figure}

Figure~\ref{evdist-daily} compares $P(\lambda)$ for {\bf \sf C}
calculated using $L=1737$ daily returns of 422 stocks for the 7-yr
period 1990--96. We find a well-defined bulk of eigenvalues that fall
within $P_{\rm rm}(\lambda)$, and deviations from $P_{\rm
rm}(\lambda)$ for large eigenvalues --- similar to what we found for
$\Delta t=30$~min [Fig.~\ref{evdist}(a)]. Thus, a comparison of
$P(\lambda)$ with the RMT result $P_{\rm rm} (\lambda)$ allows us to
distinguish the {\it bulk} of the eigenvalue spectrum of {\bf \sf C}
that agrees with RMT (random correlations) from the deviations
(genuine correlations).

\section{Universal properties: Are the bulk of eigenvalues of {\bf \sf C}
consistent with RMT?}

The presence of a well-defined bulk of eigenvalues that agree with
$P_{\rm rm}(\lambda)$ suggests that the contents of {\bf \sf C} are mostly
random except for the eigenvalues that deviate. Our conclusion was
based on the comparison of the eigenvalue distribution $P(\lambda)$ of
{\bf \sf C} with that of random matrices of the type {\bf \sf R = ${1
\over L}$ A A$^{T}$}. Quite generally, comparison of the eigenvalue
distribution with $P_{\rm rm}(\lambda)$ alone is not sufficient to
support the possibility that the bulk of the eigenvalue spectrum of
{\bf \sf C} is random. Random matrices that have drastically different
$P(\lambda)$ share similar correlation structures in their eigenvalues
--- universal properties --- that depend only on the general
symmetries of the matrix~\cite{Mehta91,Brody81,Guhr98}. Conversely,
matrices that have the same eigenvalue distribution can have
drastically different eigenvalue correlations. Therefore, a test of
randomness of {\bf \sf C} involves the investigation of correlations
in the eigenvalues $\lambda_i$.

Since by definition {\bf \sf C} is a real symmetric matrix, we shall
test the eigenvalue statistics {\bf \sf C} for universal features of
eigenvalue correlations displayed by real symmetric random matrices.
Consider a $M\times M$ real symmetric random matrix {\bf \sf S} with
off-diagonal elements $S_{ij}$, which for $i<j$ are independent and
identically distributed with zero mean $\langle S_{ij} \rangle =0$ and
variance $\langle S^2_{ij} \rangle >0$. It is conjectured based on
analytical~\cite{noteuniv} and extensive numerical
evidence~\cite{Mehta91} that in the limit $M\rightarrow \infty$,
regardless of the distribution of elements $S_{ij}$, this class of
matrices, on the scale of local mean eigenvalue spacing, display the
universal properties (eigenvalue correlation functions) of the
ensemble of matrices whose elements are distributed according to a
Gaussian probability measure --- called the Gaussian orthogonal
ensemble (GOE)~\cite{Mehta91}.

Formally, GOE is defined on the space of real symmetric matrices by
two requirements~\cite{Mehta91}. The first is that the ensemble is
invariant under orthogonal transformations, i.e., for any GOE matrix
{\bf \sf Z}, the transformation {\bf \sf Z}$\rightarrow${Z$^{\prime}
\equiv$\bf \sf W$^T$ Z W}, where {\bf \sf W} is any real orthogonal
matrix ({\bf \sf W W$^T$=I}), leaves the joint probability $P(Z) dZ$
of elements $Z_{ij}$ unchanged: $P(Z^{\prime}) dZ^{\prime} = P(Z) dZ$.
The second requirement is that the elements $\{Z_{ij}; i \leq j\}$ are
statistically independent~\cite{Mehta91}.

By definition, random cross-correlation matrices {\bf \sf R}
(Eq.~(\ref{matrixA})) that we are interested in are not strictly
GOE-type matrices, but rather belong to a special ensemble called the
``chiral'' GOE~\cite{Guhr98,Verbaarschot00}. This can be seen by the
following argument. Define a matrix {\bf \sf B}
\begin{eqnarray} \rm{\bf \sf B} \equiv \left[
\begin{array}{cc} 0 & {\bf \sf G} \\ {\bf \sf G}^T & 0 \end{array} \right] \,.
\end{eqnarray}
The eigenvalues $\gamma$ of {\bf \sf B} are given by $\det(\gamma^2
{\bf \sf I} - {\bf \sf G G}^T)=0$ and similarly, the eigenvalues
$\lambda$ of {\bf \sf R} are given by $\det(\lambda {\bf \sf I} - {\bf
\sf G G}^T)=0$. Thus, all non-zero eigenvalues of {\bf \sf B} occur in
pairs, i.e., for every eigenvalue $\lambda$ of {\bf \sf R},
$\gamma_{\pm} = \pm \sqrt{\lambda}$ are eigenvalues of {\bf \sf B}. Since the
eigenvalues occur pairwise, the eigenvalue spectra of both {\bf \sf B}
and {\bf \sf R} have special properties in the neighborhood of zero
that are different from the standard
GOE~\cite{Guhr98,Verbaarschot00}. As these special properties decay
rapidly as one goes further from zero, the eigenvalue correlations of
{\bf \sf R} in the bulk of the spectrum are still consistent with
those of the standard GOE. Therefore, our goal shall be to test the
bulk of the eigenvalue spectrum of the empirically-measured
cross-correlation matrix {\bf \sf C} with the known universal features
of standard GOE-type matrices.

In the following, we test the statistical properties of the
eigenvalues of {\bf \sf C} for three known universal
properties~\cite{Mehta91,Brody81,Guhr98} displayed by GOE matrices:
(i) the distribution of nearest-neighbor eigenvalue spacings $P_{\rm
nn} (s)$, (ii) the distribution of next-nearest-neighbor eigenvalue
spacings $P_{\rm nnn} (s)$, and (iii) the ``number variance''
statistic $\Sigma^2$.

The analytical results for the three properties listed above hold if
the spacings between adjacent eigenvalues (rank-ordered) are expressed
in units of {\it average} eigenvalue spacing. Quite generally, the
average eigenvalue spacing changes from one part of the eigenvalue
spectrum to the next. So, in order to ensure that the eigenvalue
spacing has a uniform {\it average} value throughout the spectrum, we
must find a transformation called ``unfolding,'' which maps the
eigenvalues $\lambda_i$ to new variables called ``unfolded
eigenvalues'' $\xi_i$, whose distribution is
uniform~\cite{Mehta91,Brody81,Guhr98}. Unfolding ensures that the
distances between eigenvalues are expressed in units of local mean
eigenvalue spacing~\cite{Mehta91}, and thus facilitates comparison
with theoretical results. The procedures that we use for unfolding the
eigenvalue spectrum are discussed in Appendix A.

\subsection{Distribution of nearest-neighbor eigenvalue spacings}

\begin{figure}[htb]
\narrowtext 
\centerline{
\epsfysize=0.7\columnwidth{\rotate[r]{\epsfbox{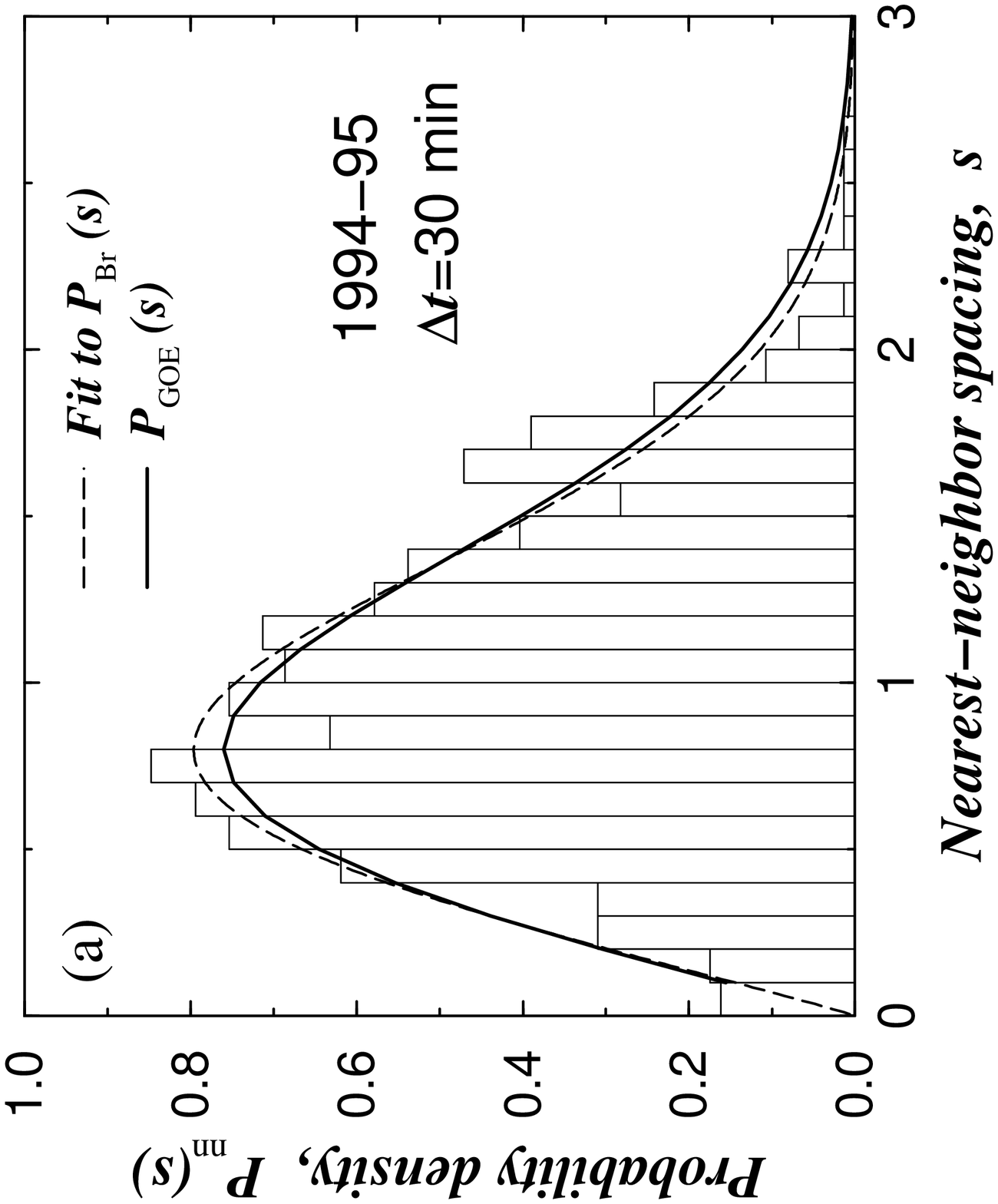}}} 
}
\vspace{0.5cm}
\centerline{
\epsfysize=0.7\columnwidth{\rotate[r]{\epsfbox{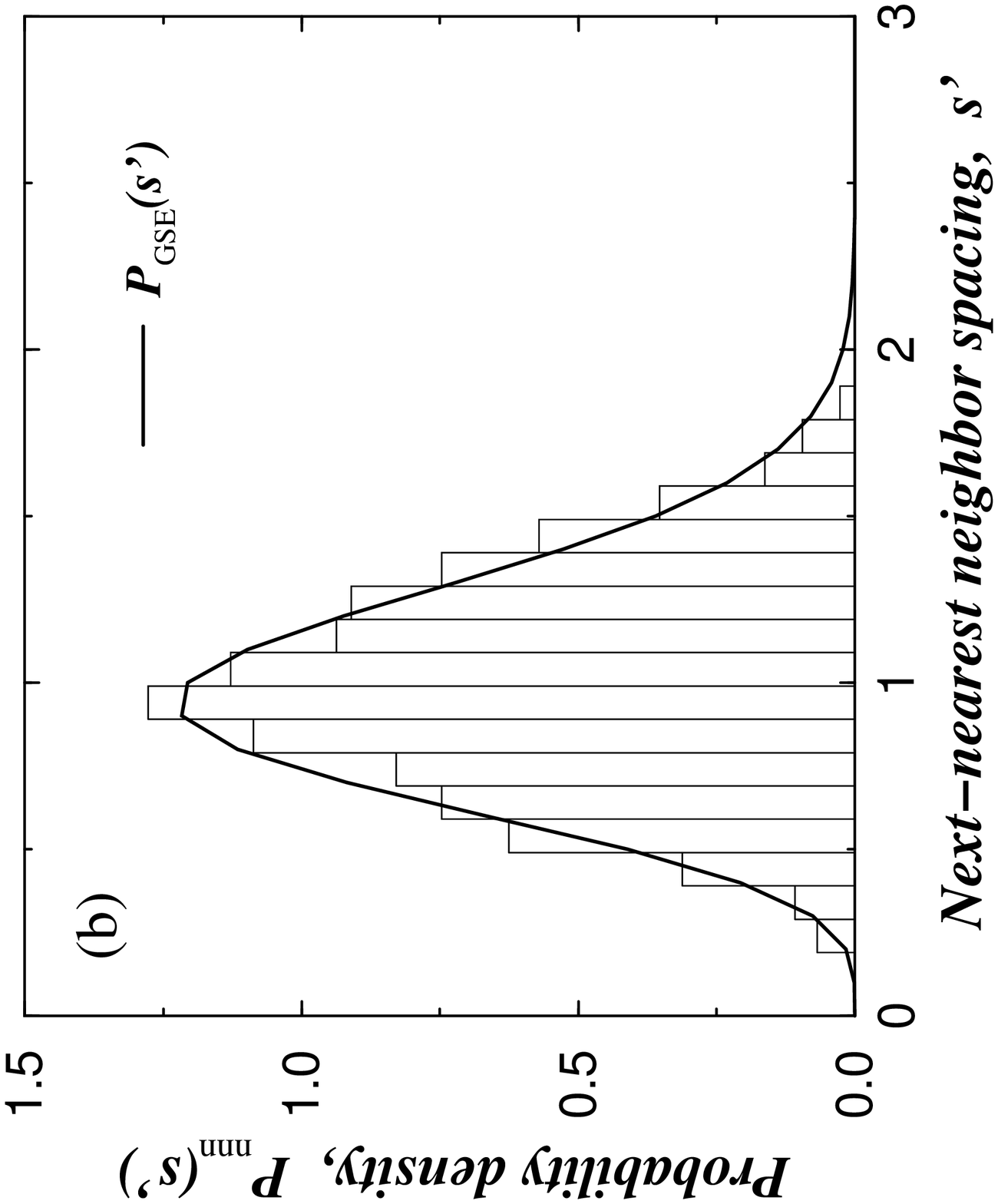}}} 
}
\caption{ (a) Nearest-neighbor ({\it nn}) spacing distribution 
$P_{\rm nn} (s)$ of the unfolded eigenvalues $\xi_i$ of {\bf \sf C}
constructed from 30-min returns for the 2-yr period 1994--95. We find
good agreement with the GOE result $P_{\rm GOE} (s)$
[Eq.~(\protect\ref{eq.3})] (solid line). The dashed line is a fit to
the one parameter Brody distribution $P_{\rm Br}$
[Eq.~(\protect\ref{defBrody})].  The fit yields $\beta = 0.99 \pm
0.02$, in good agreement with the GOE prediction $\beta=1$. A
Kolmogorov-Smirnov test shows that the GOE is $10^{5}$ times more
likely to be the correct description than the Gaussian unitary
ensemble, and $10^{20}$ times more likely than the GSE.  (b)
Next-nearest-neighbor ({\it nnn}) eigenvalue spacing distribution
$P_{\rm nnn} (s)$ of {\bf \sf C} compared to the nearest-neighbor
spacing distribution of GSE shows good agreement. A Kolmogorov-Smirnov
test cannot reject the hypothesis that $P_{\rm GSE}(s)$ is the correct
distribution at the 65\% confidence level. The results shown above are
using the Gaussian broadening procedure. Using the second procedure of
fitting $F(\lambda)$ (Appendix A) yields similar results.}
\label{spacdist}
\end{figure}

We first consider the eigenvalue spacing distribution, which reflects
two-point as well as eigenvalue correlation functions of all
orders. We compare the eigenvalue spacing distribution of {\bf \sf C}
with that of GOE random matrices. For GOE matrices, the distribution
of ``nearest-neighbor'' eigenvalue spacings $s\equiv
\xi_{k+1} -\xi_k$ is given by~\cite{Mehta91,Brody81,Guhr98}
%
\begin{equation}
P_{\rm GOE}(s)= {\pi s \over 2}\, \exp\left(- {\pi \over 4}\, s^2
\right)\,,
\label{eq.3}
\end{equation}
%
often referred to as the ``Wigner surmise''~\cite{notenn}. The
Gaussian decay of $P_{\rm GOE}(s)$ for large $s$ [bold curve in
Fig.~\ref{spacdist}(a)] implies that $P_{\rm GOE}(s)$ ``probes''
scales only of the order of one eigenvalue spacing. Thus, the spacing
distribution is known to be robust across different unfolding
procedures~\cite{Guhr98}.

We first calculate the distribution of the ``nearest-neighbor
spacings'' $s\equiv \xi_{k+1} -\xi_k$ of the unfolded eigenvalues
obtained using the Gaussian broadening
procedure. Figure~\ref{spacdist}(a) shows that the distribution
$P_{\rm nn}(s)$ of nearest-neighbor eigenvalue spacings for {\bf \sf
C} constructed from 30-min returns for the 2-yr period 1994--95 agrees
well with the RMT result $P_{\rm GOE}(s)$ for GOE matrices.

Identical results are obtained when we use the alternative unfolding
procedure of fitting the eigenvalue distribution. In addition, we test
the agreement of $P_{\rm nn}(s)$ with RMT results by fitting $P_{\rm
nn}(s)$ to the one-parameter Brody distribution~\cite{Brody81,Guhr98}
%
\begin{equation}
P_{\rm Br}(s)= B\,(1+\beta)\, s^{\beta}\,\exp(-B s^{1+\beta})\,,
\label{defBrody}
\end{equation}
%
where $B\equiv [\Gamma ({\beta+2 \over \beta+1})]^{1+\beta}$. The case
$\beta=1$ corresponds to the GOE and $\beta =0$ corresponds to
uncorrelated eigenvalues (Poisson-distributed spacings). We obtain
$\beta = 0.99 \pm 0.02$, in good agreement with the GOE prediction
$\beta=1$. To test non-parametrically that $P_{\rm GOE}(s)$ is the
correct description for $P_{\rm nn}(s)$, we perform the
Kolmogorov-Smirnov test. We find that at the 60$\%$ confidence level,
a Kolmogorov-Smirnov test cannot reject the hypothesis that the GOE is
the correct description for $P_{\rm nn}(s)$.

\begin{figure}[htb]
\narrowtext 
\centerline{
\epsfysize=1.0\columnwidth{\rotate[r]{\epsfbox{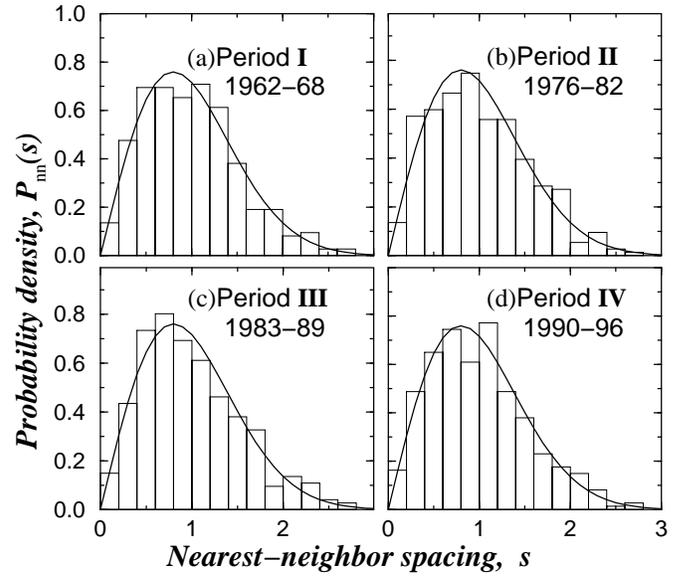}}} 
}
\vspace{0.5cm}
\caption{Nearest-neighbor spacing distribution $P(s)$ of the unfolded
eigenvalues $\xi_i$ of {\bf \sf C} computed from the daily returns of
422 stocks for the 7-yr periods (a) 1962--68 (b) 1976--82 (c)
1983--89, and (d) 1990--96.  We find good agreement with the GOE
result (solid curve). The unfolding was performed by using the
procedure of fitting the cumulative distribution of eigenvalues
(Appendix A). Gaussian broadening procedure also yields similar
results.}
\label{spacdist-daily}
\end{figure}

Next, we analyze the nearest-neighbor spacing distribution $P_{\rm
nn}(s)$ for {\bf \sf C} constructed from daily returns for four 7-yr
periods [Fig.~\ref{spacdist-daily}]. We find good agreement with the
GOE result of Eq.~(\ref{eq.3}), similar to what we find for {\bf \sf
C} constructed from 30-min returns. We also test that both of the
unfolding procedures discussed in Appendix A yield consistent results.
Thus, we have seen that the eigenvalue-spacing distribution of
empirically-measured cross-correlation matrices {\bf \sf C} is
consistent with the RMT result for real symmetric random matrices.

\subsection{Distribution of next-nearest-neighbor eigenvalue spacings}

A second independent test for GOE is the distribution $P_{\rm nnn}
(s^{\prime})$ of {\it next}-nearest-neighbor spacings $s^{\prime}
\equiv \xi_{k+2} -\xi_k$ between the unfolded eigenvalues. For 
matrices of the GOE type, according to a theorem due to
Ref.~\cite{Mehta63}, the next-nearest neighbor spacings follow the
statistics of the Gaussian symplectic ensemble
(GSE)~\cite{Mehta91,Brody81,Guhr98,notegse}. In particular, the distribution
of next-nearest-neighbor spacings $P_{\rm nnn}(s^{\prime})$ for a GOE
matrix is identical to the distribution of nearest-neighbor spacings
of the Gaussian symplectic ensemble
(GSE)~\cite{Mehta91,Guhr98}. Figure~\ref{spacdist}(b) shows that
$P_{\rm nnn}(s^{\prime})$ for the same data as Fig.~\ref{spacdist}(a)
agrees well with the RMT result for the distribution of
nearest-neighbor spacings of GSE matrices,
\begin{equation}
P_{\rm GSE}(s)= {2^{18} \over 3^6 \pi^3}\, s^4\, \exp\left(- {64 \over
9 \pi}\, s^2 \right)\,.
\label{eq.4}
\end{equation}

\subsection{Long-range eigenvalue correlations}

\begin{figure}[htb]
\narrowtext 
\centerline{
\epsfysize=0.7\columnwidth{\rotate[r]{\epsfbox{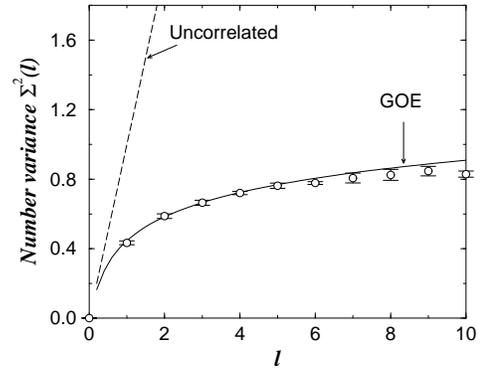}}} 
}
\caption{ (a) Number
variance $\Sigma^2(\ell)$ calculated from the unfolded eigenvalues
$\xi_i$ of {\bf \sf C } constructed from 30-min returns for the 2-yr
period 1994--95. We used Gaussian broadening procedure with the
broadening parameter $a=15$. We find good agreement with the GOE
result of Eq.~\protect\ref{defSigmagoe} (solid curve). The dashed line
corresponds to the uncorrelated case (Poisson). For the range of
$\ell$ shown, unfolding by fitting also yields similar results.}
\label{numbervar}
\end{figure}

To probe for larger scales, pair correlations (``two-point''
correlations) in the eigenvalues, we use the statistic $\Sigma^2$
often called the ``number variance,'' which is defined as the variance
of the number of unfolded eigenvalues in intervals of length $\ell$
around each $\xi_i$~\cite{Mehta91,Brody81,Guhr98},
\begin{equation}
\Sigma^2(\ell)\equiv \langle [n(\xi,\ell) -\ell]^2 \rangle_{\xi}\,,
\label{sigma2}
\end{equation}
where $n(\xi,\ell)$ is the number of unfolded eigenvalues in the interval
$[\xi-\ell/2,\xi+\ell/2]$ and $\langle\dots\rangle_{\xi}$ denotes an average
over all $\xi$. If the eigenvalues are uncorrelated, $\Sigma^2 \sim
\ell$. For the opposite extreme of a ``rigid'' eigenvalue spectrum
(e.g. simple harmonic oscillator), $\Sigma^2$ is a constant. Quite
generally, the number variance $\Sigma^2$ can be expressed as
\begin{equation}
\Sigma^2(\ell)=\ell-2\,\int_0^\ell\,(\ell-x) \, Y(x) dx\,,
\label{defSigmagoe}
\end{equation}
where $Y(x)$ (called ``two-level cluster function'') is related to the
two-point correlation function [c.f., Ref.~\cite{Mehta91}, pp.79]. For
the GOE case, $Y(x)$ is explicitly given by
\begin{equation}
Y(x)\equiv s^2(x)+{ds\over dx}\,\int_{x}^{\infty} s(x^{\prime}) dx^{\prime}\,,
\label{defy2}
\end{equation}
where
\begin{equation}
s(x) \equiv {\sin(\pi x) \over \pi x}\,.
\label{defs}
\end{equation}
For large values of $\ell$, the number variance $\Sigma^2$ for GOE has
the ``intermediate'' behavior 
\begin{equation}
\Sigma^2 \sim \ln \ell.
\label{sigmalog}
\end{equation}
Figure~\ref{numbervar} shows that $\Sigma^2(\ell)$ for {\bf \sf C}
calculated using 30-min returns for 1994--95 agrees well with the RMT
result of Eq.~(\ref{defSigmagoe}). For the range of $\ell$ shown in
Fig.~\ref{numbervar}, both unfolding procedures yield similar results.
Consistent results are obtained for {\bf \sf C} constructed from daily
returns.

\subsection{Implications}

To summarize this section, we have tested the statistics of {\bf \sf
C} for universal features of eigenvalue correlations displayed by GOE
matrices. We have seen that the distribution of the nearest-neighbor
spacings $P_{\rm nn}(s)$ is in good agreement with the GOE result. To
test whether the eigenvalues of {\bf \sf C} display the RMT results
for long-range two-point eigenvalue correlations, we analyzed the
number variance $\Sigma^2$ and found good agreement with GOE
results. Moreover, we also find that the statistics of next-nearest
neighbor spacings conform to the predictions of RMT. These findings
show that the statistics of the {\it bulk} of the eigenvalues of the
empirical cross-correlation matrix {\bf \sf C} is consistent with
those of a real symmetric random matrix. Thus, information about
genuine correlations are contained in the deviations from RMT, which
we analyze below.

\section{Statistics of eigenvectors}

\subsection{Distribution of eigenvector components}

The deviations of $P(\lambda)$ from the RMT result $P_{\rm
rm}(\lambda)$ suggests that these deviations should also be displayed
in the statistics of the corresponding eigenvector
components~\cite{Laloux99}. Accordingly, in this section, we analyze the distribution
of eigenvector components. The distribution of the components
$\{u^k_{l}; l = 1,\dots,N\}$ of eigenvector {\bf \sf u$^k$} of a
random correlation matrix {\bf \sf R} should conform to a Gaussian
distribution with mean zero and unit variance~\cite{Guhr98},
\begin{equation}
\rho_{\rm rm} (u) = {1 \over \sqrt{2 \pi}} \exp({-u^2 \over 2})\,.
\label{port-thomas}
\end{equation}

\begin{figure}[h]
\narrowtext 
\centerline{
\epsfysize=1.0\columnwidth{\rotate[r]{\epsfbox{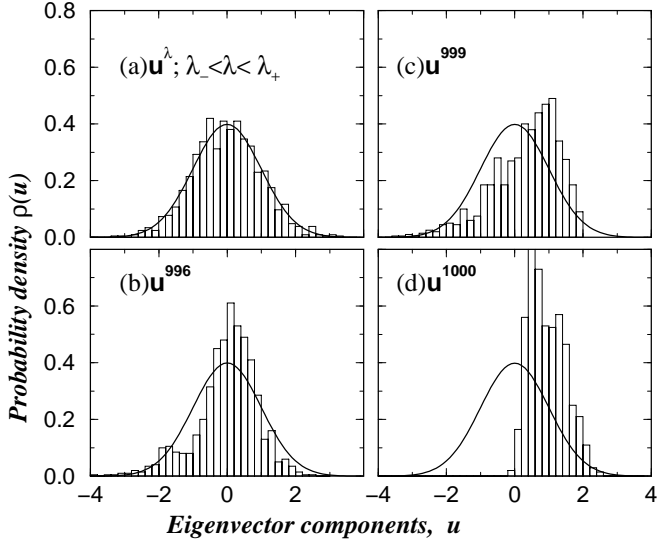}}} 
}
\vspace{0.5cm}
\caption{ (a) Distribution $\rho(u)$ of eigenvector components for 
one eigenvalue in the bulk $\lambda_-<\lambda<\lambda_+$ shows good
agreement with the RMT prediction of Eq.~(\protect\ref{port-thomas})
(solid curve).  Similar results are obtained for other eigenvalues in
the bulk.  $\rho(u)$ for (b) {\bf \sf u$^{996}$} and (c) {\bf \sf
u$^{999}$}, corresponding to eigenvalues larger than the RMT upper
bound $\lambda_+$ (shaded region in Fig.~\protect\ref{evdist}). (d)
$\rho(u)$ for {\bf \sf u$^{1000}$} deviates significantly from the
Gaussian prediction of RMT. The above plots are for {\bf \sf C}
constructed from 30-min returns for the 2-yr period 1994--95. We also
obtain similar results for {\bf \sf C} constructed from daily
returns.}
\label{distevec}
\end{figure}

First, we compare the distribution of eigenvector components of {\bf
\sf C} with Eq.~(\ref{port-thomas}).  We analyze $\rho(u)$ for 
{\bf \sf C} computed using 30-min returns for 1994--95. We choose one
typical eigenvalue $\lambda_k$ from the bulk ($\lambda_- \leq \lambda_k
\leq \lambda_+$) defined by $P_{\rm rm}(\lambda)$ of Eq.~(\ref{densuncorr}). 
Figure~\ref{distevec}(a) shows that $\rho(u)$ for a typical {\bf \sf
u$^k$} from the bulk shows good agreement with the RMT result
$\rho_{\rm rm}(u)$. Similar analysis on the other eigenvectors
belonging to eigenvalues within the bulk yields consistent results, in
agreement with the results of the previous sections that the bulk
agrees with random matrix predictions. We test the agreement of the
distribution $\rho(u)$ with $\rho_{\rm rm}(u)$ by calculating the
kurtosis, which for a Gaussian has the value $3$. We find significant
deviations from $\rho_{\rm rm}(u)$ for $\approx 20$ largest and
smallest eigenvalues. The remaining eigenvectors have values of
kurtosis that are consistent with the Gaussian value $3$.

Consider next the ``deviating'' eigenvalues $\lambda_i$, larger than
the RMT upper bound, $\lambda_i >
\lambda_{+}$. Figure~\ref{distevec}(b) and (c) show that, for deviating 
eigenvalues, the distribution of eigenvector components $\rho(u)$
deviates systematically from the RMT result $\rho_{\rm rm}(u)$.
Finally, we examine the distribution of the components of the
eigenvector {\bf
\sf u$^{1000}$} corresponding to the largest eigenvalue
$\lambda_{1000}$. Figure~\ref{distevec}(d) shows that $\rho(u^{1000})$
deviates remarkably from a Gaussian, and is approximately uniform,
suggesting that all stocks participate. In addition, we find that
almost all components of {\bf \sf u$^{1000}$} have the same sign, thus
causing $\rho(u)$ to shift to one side. This suggests that the
significant participants of eigenvector {\bf \sf u$^k$} have a common
component that affects all of them with the same bias.

\subsection{Interpretation of the largest eigenvalue and the corresponding
eigenvector}

\begin{figure}[hbt]
\narrowtext 
\centerline{
\epsfysize=0.7\columnwidth{\rotate[r]{\epsfbox{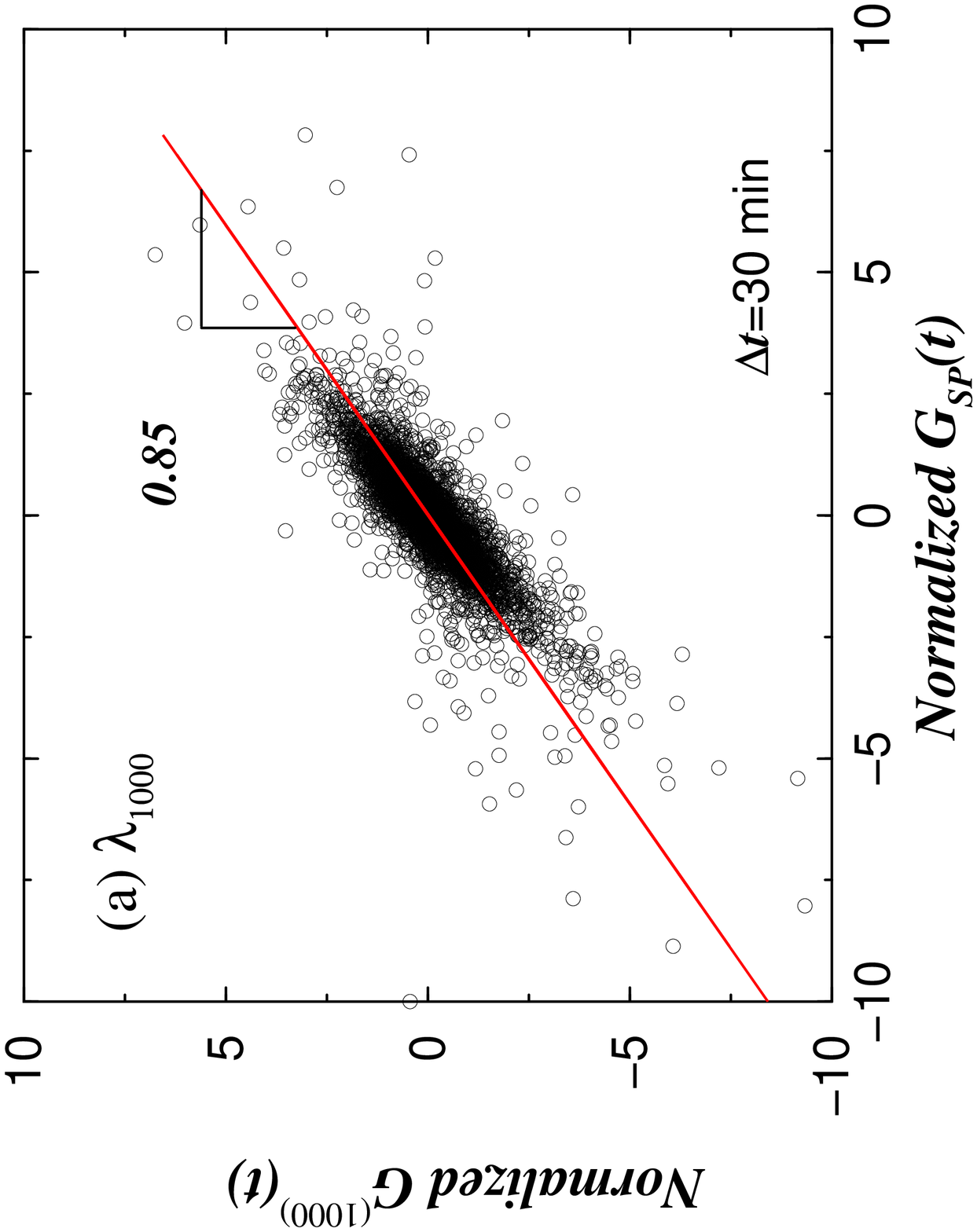}}} 
}
\centerline{
\epsfysize=0.7\columnwidth{\rotate[r]{\epsfbox{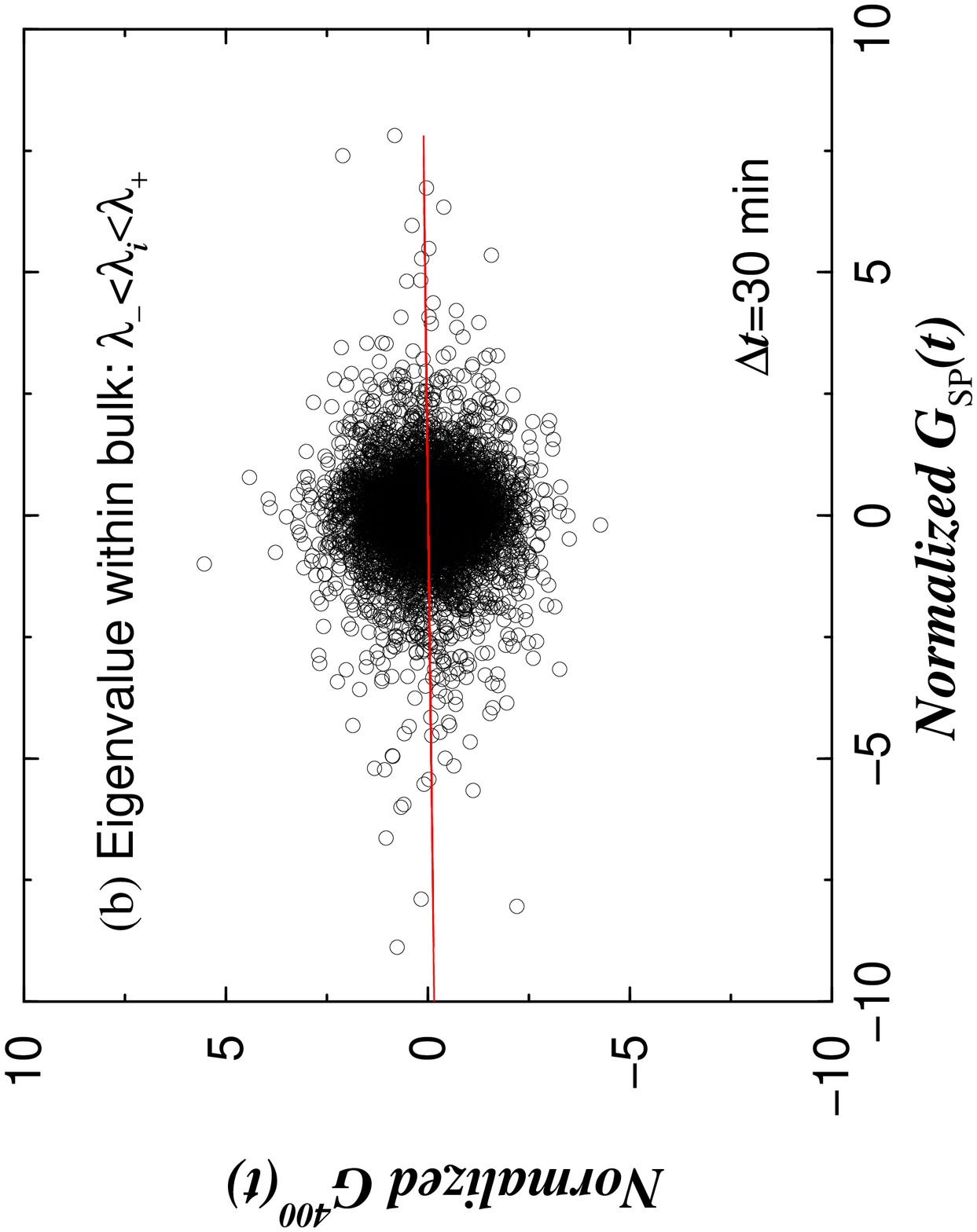}}} 
}
\caption{(a) S\&P 500 returns at $\Delta t=30$~min regressed against 
the 30-min return on the portfolio $G^{1000}$
(Eq.~(\protect\ref{marketp})) defined by the eigenvector {\bf \sf
u$^{1000}$}, for the 2-yr period 1994--95.  Both axes are scaled by
their respective standard deviations. A linear regression yields a
slope $0.85 \pm 0.09$. (b) Return (in units of standard deviations) on
the portfolio defined by an eigenvector corresponding to an eigenvalue
$\lambda_{400}$ within the RMT bounds regressed against the normalized
returns of the S\&P 500 index shows no significant dependence.  Both
axes are scaled by their respective standard deviations. The slope of
the linear fit is $0.014\pm 0.011$, close to $0$ indicating that the
dependence between $G^{1000}$ and $G_{\rm SP} (t)$ found in part (a)
is statistically significant.}
\label{sp-market}
\end{figure}

Since all components participate in the eigenvector corresponding to
the largest eigenvalue, it represents an influence that is common to
all stocks. Thus, the largest eigenvector quantifies the qualitative
notion that certain newsbreaks (e.g., an interest rate increase)
affect all stocks alike~\cite{CLM}. One can also interpret the largest
eigenvalue and its corresponding eigenvector as the collective
`response' of the entire market to stimuli. We quantitatively
investigate this notion by comparing the projection (scalar product)
of the time series {\bf \sf G} on the eigenvector {\bf \sf
u$^{1000}$}, with a standard measure of US stock market performance
--- the returns $G_{\rm SP}(t)$ of the S\&P 500 index. We calculate
the projection $G^{1000}(t)$ of the time series $G_j(t)$ on the
eigenvector {\bf \sf u$^{1000}$},
\begin{equation}
G^{1000}(t)\equiv \sum_{j=1}^{1000} u^{1000}_{j}\, G_{j}(t)\,.
\label{marketp}
\end{equation}
By definition, $G^{1000}(t)$ shows the return of the portfolio defined
by {\bf \sf u$^{1000}$}. We compare $G^{1000}(t)$ with $G_{\rm
SP}(t)$, and find remarkably similar behavior for the two, indicated
by a large value of the correlation coefficient $\langle G_{\rm SP}(t)
G^{1000}(t) \rangle = 0.85$. Figure~\ref{sp-market} shows
$G^{1000}(t)$ regressed against $G_{SP}(t)$, which shows relatively
narrow scatter around a linear fit. Thus, we interpret the eigenvector
{\bf \sf u$^{1000}$} as quantifying market-wide influences on all
stocks~\cite{Laloux99,Plerou99}.

We analyze {\bf \sf C} at larger time scales of $\Delta t=1$~day and
find similar results as above, suggesting that similar correlation
structures exist for quite different time scales. Our results for the
distribution of eigenvector components agree with those reported in
Ref.~\cite{Laloux99}, where $\Delta t= 1$~day returns are analyzed.
We next investigate how the largest eigenvalue changes as a function
of time.  Figure~\ref{volat} shows the time dependence~\cite{Drodz99}
of the largest eigenvalue ($\lambda_{422}$) for the 35-yr period
1962--96. We find large values of the largest eigenvalue during
periods of high market volatility, which suggests strong collective
behavior in regimes of high volatility.

One way of statistically modeling an influence that is common to all
stocks is to express the return $G_i$ of stock $i$ as
\begin{equation}
G_i(t) = \alpha_i + \beta_i M(t) + \epsilon_i(t)\,,
\label{marketm}
\end{equation}
where $M(t)$ is an additive term that is the same for all stocks,
$\langle \epsilon (t) \rangle = 0$, $\alpha_i$ and $\beta_i$ are
stock-specific constants, and $\langle M(t) \epsilon(t)
\rangle = 0$. This common term $M(t)$ gives rise to correlations between 
any pair of stocks. The decomposition of Eq.~(\ref{marketm}) forms the
basis of widely-used economic models, such as multi-factor models and
the Capital Asset Pricing
Model~\cite{CLM,Sharpe70,Sharpe95,Sharpe64,Lintner65,Ross76,Brown85,Black72,Blume73,Fama92,FamaMacbeth73,RollRoss94,Chen86,Merton73a,Lehmann88,Campbell96,Campbell93a,Conner86a}.
Since {\bf \sf u$^{1000}$} represents an influence that is common to
all stocks, we can approximate the term $M(t)$ with $G^{1000}(t)$. The
parameters $\alpha_i$ and $\beta_i$ can therefore be estimated by an
ordinary least squares regression.

\begin{figure}[hbt]
\narrowtext 
\centerline{
\epsfysize=0.7\columnwidth{\rotate[r]{\epsfbox{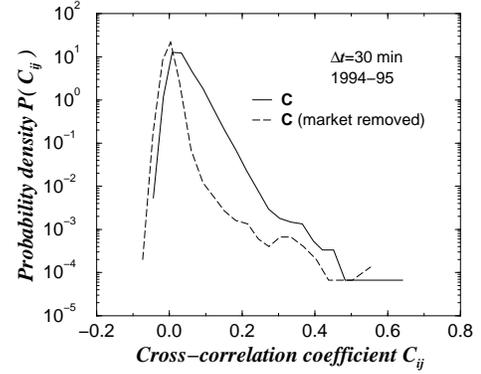}}}
}
\caption{Probability distribution $P(C_{ij})$ of the cross-correlation
coefficients for the 2-yr period 1994--95 before and after removing the
effect of the largest eigenvalue $\lambda_{1000}$.  Note that removing
the effect of $\lambda_{1000}$ shifts $P(C_{ij})$ toward a smaller
average value $\langle C_{ij} \rangle =0.002$ compared to the original
value $\langle C_{ij} \rangle =0.03$.}
\label{corr-ev1000}
\end{figure}

Next, we remove the contribution of $G^{1000}(t)$ to each time series
$G_i(t)$, and construct {\bf \sf C} from the residuals $\epsilon_i(t)$
of Eq.~(\ref{marketm}). Figure~\ref{corr-ev1000} shows that the
distribution $P(C_{ij})$ thus obtained has significantly smaller
average value $\langle C_{ij} \rangle$, showing that a large degree of
cross-correlations contained in {\bf \sf C} can be attributed to the
influence of the largest eigenvalue (and its corresponding
eigenvector)~\cite{Cizeau00,Lillo00a}.

\subsection{Number of significant participants in an eigenvector: Inverse
Participation Ratio}

Having studied the interpretation of the largest eigenvalue which
deviates significantly from RMT results, we next focus on the
remaining eigenvalues. The deviations of the distribution of
components of an eigenvector {\bf \sf u$^k$} from the RMT prediction
of a Gaussian is more pronounced as the separation from the RMT upper
bound $\lambda_k - \lambda_+$ increases. Since proximity to
$\lambda_+$ increases the effects of randomness, we quantify the
number of components that participate significantly in each
eigenvector, which in turn reflects the degree of deviation from RMT
result for the distribution of eigenvector components. To this end, we
use the notion of the inverse participation ratio (IPR), often applied
in localization theory~\cite{Guhr98,bandmatrix}. The IPR of the
eigenvector {\bf \sf u$^k$} is defined as
\begin{equation}
I^k \equiv \sum_{l=1}^N\, [u^k_{l}]~^4\,,
\label{defIPR}
\end{equation}
where $u^k_{l}$, $l=1,\dots,1000$ are the components of
eigenvector {\bf \sf u$^k$}. The meaning of $I^k$ can be illustrated
by two limiting cases: (i) a vector with identical components
$u^k_{l}\equiv 1/\sqrt{N}$ has $I^k=1/N$, whereas (ii) a vector
with one component $u^k_1=1$ and the remainder zero has $I^k=1$. Thus,
the IPR quantifies the reciprocal of the number of eigenvector
components that contribute significantly.

\begin{figure}[h]
\narrowtext 
\centerline{
\epsfysize=1.0\columnwidth{\rotate[r]{\epsfbox{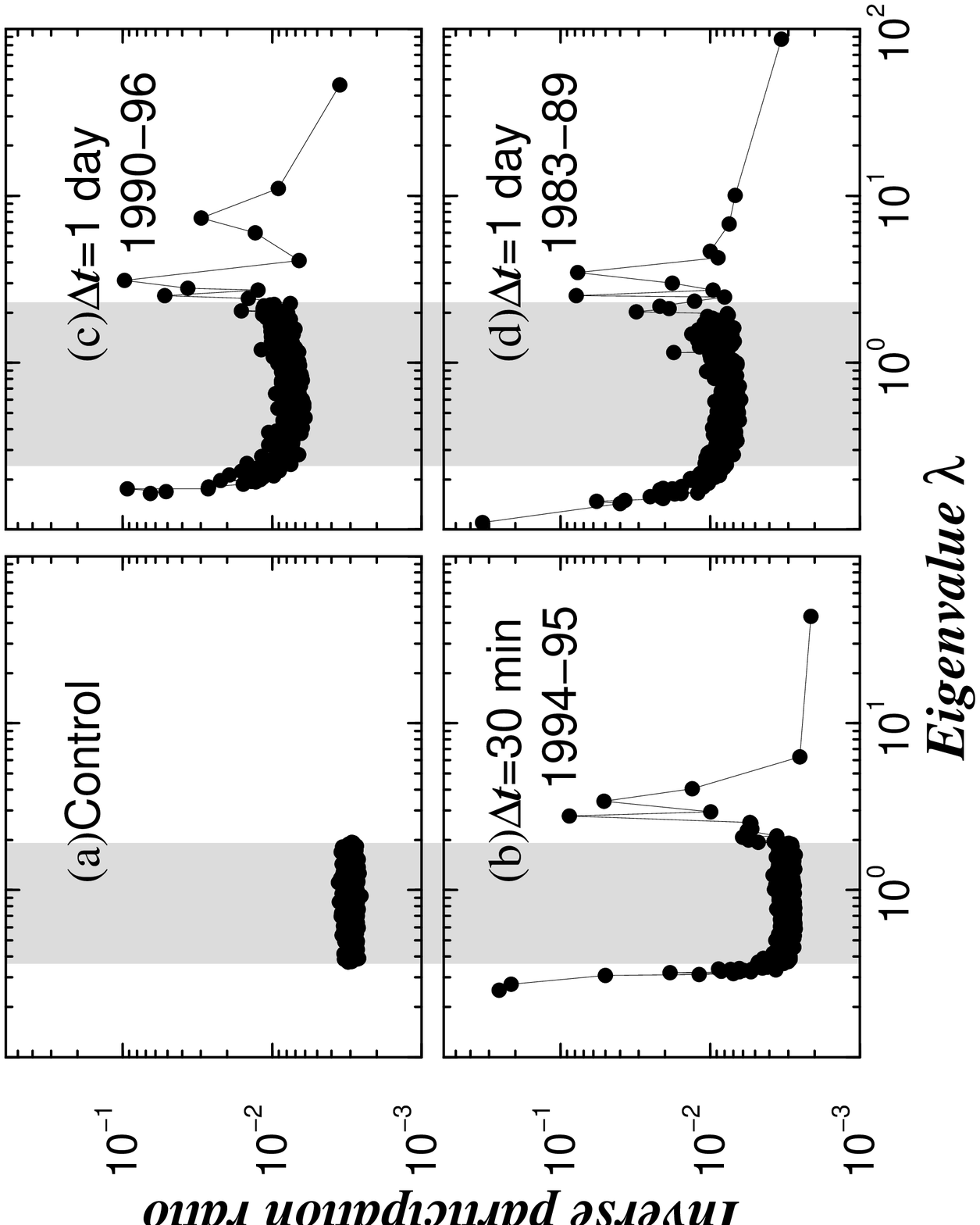}}} 
}
\vspace{1cm}
\caption{ (a) Inverse participation ratio (IPR) as a function of 
eigenvalue $\lambda$ for the random cross-correlation matrix {\bf \sf
R} of Eq.~(\protect\ref{densuncorr}) constructed using $N=1000$
mutually uncorrelated time series of length $L=6448$. IPR for {\bf \sf
C} constructed from (b) 6448 records of 30-min returns for 1000 stocks
for the 2-yr period 1994--95, (c) 1737 records of 1-day returns for
422 stocks in the 7-yr period 1990--96, and (d) 1737 records of 1-day
returns for 422 stocks in the 7-yr period 1983--89. The shaded regions
show the RMT bounds $[\lambda_+,\lambda_-]$.}
\label{ipr}
\end{figure}

Figure~\ref{ipr}(a) shows $I^k$ for the case of the control of
Eq.~(\ref{matrixA}) using time series with the empirically-found
distribution of returns~\cite{Plerou99b}.  The average value of $I^k$
is $\langle I \rangle \approx 3\times 10^{-3} \approx 1/N$ with a
narrow spread, indicating that the vectors are {\it
extended\/}~\cite{bandmatrix,extended}---i.e., almost all components
contribute to them. Fluctuations around this average value are
confined to a narrow range (standard deviation of $1.5 \times
10^{-4}$).  

Figure~\ref{ipr}(b) shows that $I^k$ for {\bf \sf C} constructed from
30-min returns from the period 1994--95, agrees with $I^k$ of the
random control in the bulk ($\lambda_- <
\lambda_i < \lambda_+$). In contrast, the edges of the eigenvalue 
spectrum of {\bf \sf C} show significant deviations of $I^k$ from
$\langle I \rangle$. The largest eigenvalue has $1/I^k \approx 600$
for the 30-min data [Fig.~\ref{ipr}(b)] and $1/I^k \approx 320$ for
the 1-day data [Fig.~\ref{ipr}(c) and (d)], showing that almost all
stocks participate in the largest eigenvector. For the rest of the
large eigenvalues which deviate from the RMT upper bound, $I^k$ values
are approximately 4-5 times larger than $\langle I \rangle$, showing
that there are varying numbers of stocks contributing to these
eigenvectors. In addition, we also find that there are large $I^k$
values for vectors corresponding to few of the small eigenvalues
$\lambda_i \approx 0.25 < \lambda_-$. The deviations at both edges of
the eigenvalue spectrum are considerably larger than $\langle I
\rangle$, which suggests that the vectors are {\it
localized\/}~\cite{bandmatrix,extended}---i.e., only a few stocks
contribute to them.

The presence of vectors with large values of $I^k$ also arises in the
theory of Anderson localization\cite{electrons}.  In the context of
localization theory, one frequently finds ``random band
matrices''\cite{bandmatrix} containing extended states with small
$I^k$ in the bulk of the eigenvalue spectrum, whereas edge states are
localized and have large $I^k$.  Our finding of localized states for
small and large eigenvalues of the cross-correlation matrix {\bf
\sf C} is reminiscent of Anderson localization and suggests that {\bf
\sf C} may have a random band matrix structure. A random band matrix 
{\bf \sf B} has elements $B_{ij}$ independently drawn from different
probability distributions. These distributions are often taken to be
Gaussian parameterized by their variance, which depends on $i$ and
$j$.  Although such matrices are random, they still contain
probabilistic information arising from the fact that a metric can be
defined on their set of indices $i$. A related, but distinct way of
analyzing cross-correlations by defining `ultra-metric' distances has
been studied in Ref.~\cite{ultra}.

\subsection{Interpretation of deviating eigenvectors 
{\bf \sf u$^{990}$}--{\bf \sf u$^{999}$}}

We quantify the number of significant participants of an eigenvector
using the IPR, and we examine the $1/I^k$ components of eigenvector
{\bf \sf u$^k$} for common features~\cite{Gopi00}. A direct
examination of these eigenvectors, however, does not yield a
straightforward interpretation of their economic relevance. To
interpret their meaning, we note that the largest eigenvalue is an
order of magnitude larger than the others, which constrains the
remaining $N-1$ eigenvalues since {\rm Tr {\bf \sf C}}~$=N$.  Thus, in
order to analyze the deviating eigenvectors, we must remove the effect
of the largest eigenvalue $\lambda_{1000}$.

\begin{figure}[h]
\narrowtext 
\centerline{
\epsfysize=0.7\columnwidth{\rotate[r]{\epsfbox{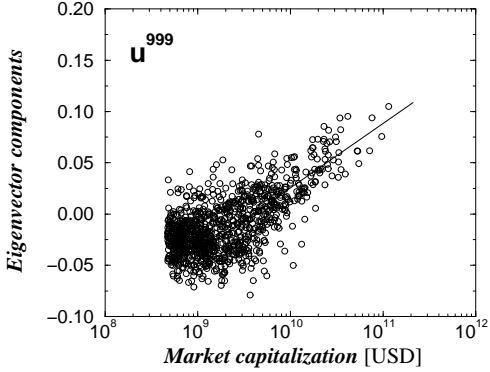}}} 
}
\caption{All $10^3$ eigenvector components of {\bf \sf u$^{999}$}
plotted against market capitalization (in units of US Dollars) shows
that firms with large market capitalization contribute
significantly. The straight line, which shows a logarithmic fit, is a
guide to the eye.}
\label{figmc}
\end{figure}

In order to avoid the effect of $\lambda_{1000}$, and thus
$G^{1000}(t)$, on the returns of each stock $G_i(t)$, we perform the
regression of Eq.~(\ref{marketm}), and compute the residuals
$\epsilon_i(t)$. We then calculate the correlation matrix {\bf \sf C}
using $\epsilon_i(t)$ in Eq.(~\ref{norm-ret}) and
Eq.~(\ref{eq.2}). Next, we compute the eigenvectors {\bf \sf u$^k$} of
{\bf \sf C} thus obtained, and analyze their significant
participants. The eigenvector {\bf \sf u$^{999}$} contains
approximately $1/I^{999} = 300$ significant participants, which are
all stocks with large values of market capitalization.
Figure~\ref{figmc} shows that the magnitude of the eigenvector
components of {\bf \sf u$^{999}$} shows an approximately logarithmic
dependence on the market capitalizations of the corresponding stocks.

\begin{figure}[hbt]
\narrowtext 
\centerline{
\epsfysize=0.8\columnwidth{\rotate[r]{\epsfbox{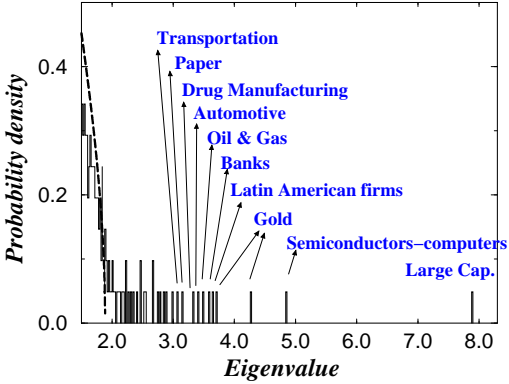}}}
}
\caption{Schematic illustration of the interpretation of the 
eigenvectors corresponding to the eigenvalues that deviate from the
RMT upper bound. The dashed curve shows the RMT result of
Eq.~(\protect\ref{densuncorr}).}
\label{evindustry}
\end{figure}

We next analyze the significant contributors of the rest of the
eigenvectors. We find that each of these deviating eigenvectors
contains stocks belonging to similar or related industries as
significant contributors. Table~\ref{industry.tab} shows the ticker
symbols and industry groups (Standard Industry Classification (SIC)
code) for stocks corresponding to the ten largest eigenvector
components of each eigenvector. We find that these eigenvectors
partition the set of all stocks into distinct groups which contain
stocks with large market capitalization ({\bf \sf u$^{999}$}), stocks
of firms in the electronics and computer industry ({\bf \sf
u$^{998}$}), a combination of gold mining and investment firms ({\bf
\sf u$^{996}$} and {\bf \sf u$^{997}$}), banking firms ({\bf \sf
u$^{994}$}), oil and gas refining and equipment ({\bf \sf u$^{993}$}),
auto manufacturing firms ({\bf \sf u$^{992}$}), drug manufacturing
firms ({\bf \sf u$^{991}$}), and paper manufacturing ({\bf \sf
u$^{990}$}).  One eigenvector ({\bf \sf u$^{995}$}) displays a mixture
of three industry groups --- telecommunications, metal mining, and
banking. An examination of these firms shows significant business
activity in Latin America.  Our results are also represented
schematically in Fig.~\ref{evindustry}. A similar classification of
stocks into sectors using different methods is obtained in
Ref.~\cite{ultra}.

Instead of performing the regression of Eq(~\ref{marketm}), one can
remove the U-shaped intra-daily pattern using the procedure of
Ref~\cite{Liu99} and compute {\bf \sf C}. The results thus obtained
are consistent with those obtained using the procedure of using the
residuals of the regression of Eq.~(\ref{marketm}) to compute {\bf \sf
C} (Table~\ref{industry.tab}). Often {\bf \sf C} is constructed from
returns at longer time scales of $\Delta t=1$~week or 1~month to avoid
short time scale effects~\cite{Elton95}.

\subsection{Smallest eigenvalues and their corresponding eigenvectors}

Having examined the largest eigenvalues, we next focus on the smallest
eigenvalues which show large values of $I^k$ [Fig.~\ref{ipr}]. We find
that the eigenvectors corresponding to the smallest eigenvalues
contain as significant participants, pairs of stocks which have the
largest values of $C_{ij}$ in our sample. For example, the two largest
components of {\bf \sf u$^1$} correspond to the stocks of Texas
Instruments (TXN) and Micron Technology (MU) with $C_{ij}=0.64$, the
largest correlation coefficient in our sample. The largest components
of {\bf \sf u$^2$} are Telefonos de Mexico (TMX) and Grupo Televisa
(TV) with $C_{ij}=0.59$ (second largest correlation coefficient). The
eigenvector {\bf \sf u$^3$} shows Newmont Gold Company (NGC) and
Newmont Mining Corporation (NEM) with $C_{ij}=0.50$ (third largest
correlation coefficient) as largest components.  In all three
eigenvectors, the relative sign of the two largest components is {\it
negative}. Thus pairs of stocks with a correlation coefficient much
larger than the average $\langle C_{ij} \rangle$ effectively
``decouple'' from other stocks.

The appearance of strongly correlated pairs of stocks in the
eigenvectors corresponding to the smallest eigenvalues of {\bf \sf C}
can be qualitatively understood by considering the example of a
$2\times 2$ cross-correlation matrix
\begin{eqnarray} \rm{\bf \sf C}_{2\times 2} = \left[
\begin{array}{cc} 1 & c \\ c & 1 \end{array} \right] \ \ .
\end{eqnarray}
The eigenvalues of {\bf \sf C}$_{2\times 2}$ are $\beta_\pm = 1 \pm
c$. The smaller eigenvalue $\beta_-$ decreases monotonically with
increasing cross-correlation coefficient $c$. The corresponding
eigenvector is the anti-symmetric linear combination of the basis
vectors $\left( \begin{array}{c} 1 \\ 0 \end{array} \right)$ and
$\left( \begin{array}{c} 0 \\ 1 \end{array} \right)$, in agreement
with our empirical finding that the relative sign of largest
components of eigenvectors corresponding to the smallest eigenvalues
is negative. In this simple example, the symmetric linear combination
of the two basis vectors appears as the eigenvector of the large
eigenvalue $\beta_+$. Indeed, we find that TXN and MU are the largest
components of {\bf \sf u$^{998}$}, TMX and TV are the largest
components of {\bf \sf u$^{995}$}, and NEM and NGC are the largest and
third largest components of {\bf \sf u$^{997}$}.


\section{Stability of eigenvectors in time}

We next investigate the degree of stability in time of the
eigenvectors corresponding to the eigenvalues that deviate from RMT
results. Since deviations from RMT results imply genuine correlations
which remain stable in the period used to compute {\bf \sf C}, we
expect the deviating eigenvectors to show some degree of time
stability.

We first identify the $p$ eigenvectors corresponding to the $p$
largest eigenvalues which deviate from the RMT upper bound
$\lambda_{+}$. We then construct a $p\times N$ matrix {\bf \sf D} with
elements $D_{kj}= \{ u^k_j\,;k=1,\dots,p\,;j=1,\dots,N\}$. Next, we
compute a $p \times p$ ``overlap matrix'' {\bf \sf O($t, \tau$) =
D$_{A}$ D$_B^{T}$}, with elements $O_{ij}$ defined as the scalar
product of eigenvector {\bf \sf u$^i$} of period A (starting at time
$t=t$) with {\bf \sf u$^j$} of period B at a later time $t+\tau$,
\begin{equation}
O_{ij}(t,\tau)\equiv \sum_{k=1}^{N} D_{ik}(t) D_{jk}(t+\tau)\,.
\label{overlap}
\end{equation}
If all the $p$ eigenvectors are ``perfectly'' non-random and stable in
time $O_{ij} = \delta_{ij}$.

We study the overlap matrices {\bf \sf O} using both high-frequency
and daily data.  For high-frequency data ($L=6448$ records at 30-min
intervals), we use a moving window of length $L=1612$, and slide it
through the entire 2-yr period using discrete time steps $L/4=403$. We
first identify the eigenvectors of the correlation matrices for each
of these time periods. We then calculate overlap matrices {\bf \sf
O($t=0,\tau = n L/4$)}, where $n\in \{1,2,3,\dots\}$, between the
eigenvectors for $t=0$ and for $t=\tau$.

\begin{figure}[hbt]
\narrowtext 
\centerline{
\epsfysize=1.0\columnwidth{\epsfbox{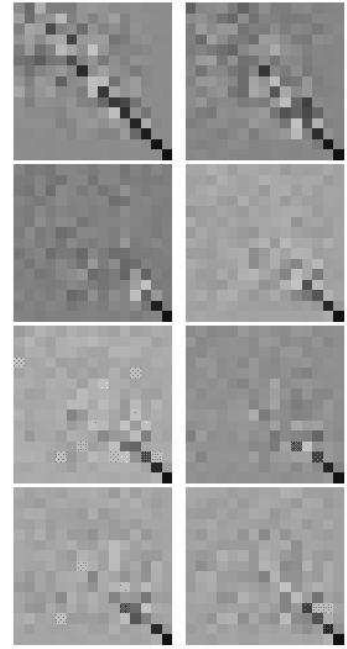}}
}
\vspace{0.05cm}
\caption{Grey scale pixel representation of the overlap matrix 
{\bf \sf O}$(t,\tau)$ as a function of time for 30-min data for the
2-yr period 1994--95.  Here, the grey scale coding is such that black
corresponds to $O_{ij}=1$ and white corresponds to $O_{ij}=0$. The
length of the time window used to compute {\bf \sf C} is $L=1612$
($\approx$60~days) and the separation $\tau=L/4=403$ used to calculate
successive $O_{ij}$. Thus, the left figure on the first row
corresponds to the overlap between the eigenvector from the starting
$t=0$ window and the eigenvector from time window $\tau=L/4$
later. The right figure is for $\tau=2L/4$. In the same way, the left
figure on the second row is for $\tau=3L/4$, the right figure for
$\tau=4L/4$, and so on. Even for large $\tau \approx 1$~yr, the
largest four eigenvectors show large values of $O_{ij}$.}
\label{overlap-taq}
\end{figure}

Figure~\ref{overlap-taq} shows a grey scale pixel-representation of
the matrix {\bf \sf O~($t,\tau$)}, for different $\tau$.  First, we
note that the eigenvectors that deviate from RMT bounds show varying
degrees of stability ($O_{ij}(t,\tau)$) in time. In particular, the
stability in time is largest for {\bf \sf u$^{1000}$}. Even at lags of
$\tau = 1$~yr the corresponding overlap $\approx 0.85$. The remaining
eigenvectors show decreasing amounts of stability as the RMT upper
bound $\lambda_+$ is approached. In particular, the 3-4 largest
eigenvectors show large values of $O_{ij}$ for up to $\tau=1$~yr.

\begin{figure}[hbt]
\narrowtext 
\centerline{
\epsfysize=1.0\columnwidth{\epsfbox{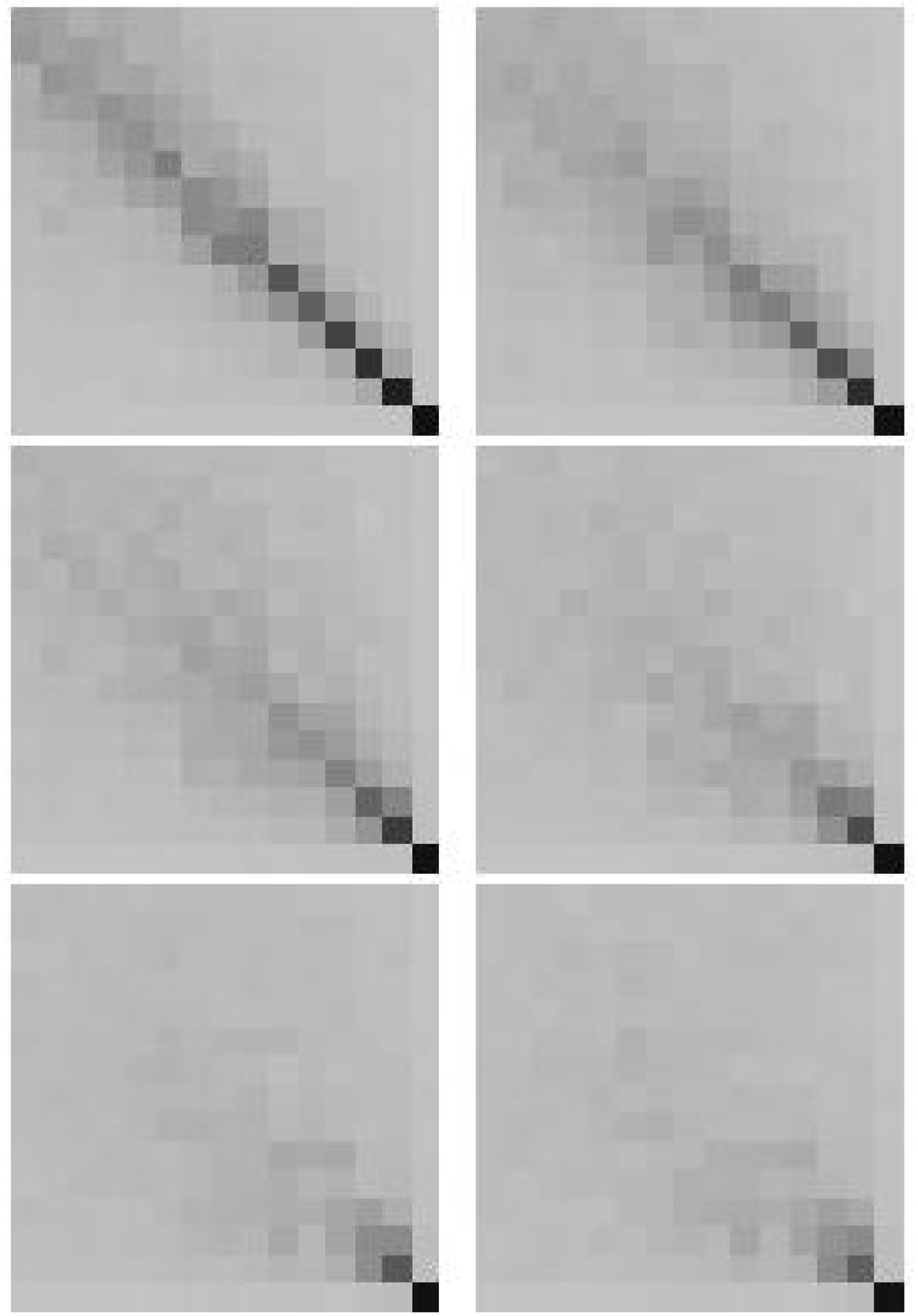}}
}
\vspace{0.05cm}
\caption{Grey scale pixel representation of the overlap matrix 
$\langle${\bf \sf O}$(t,\tau)\rangle_t$ for 1-day data, where we have
averaged over all starting points $t$. Here, the length of the time
window used to compute {\bf \sf C} is $L=965$ ($\approx$4~yr) and the
separation $\tau=L/5=193$~days used to calculate $O_{ij}$. Thus, the
left figure on the first row is for $\tau=L/5$ and the right figure is
for $\tau=2L/5$. In the same way, the left figure on the second row is
for $\tau=3L/5$, the right figure for $\tau=4L/5$, and so on. Even for
large $\tau \approx 20$~yr, the largest two eigenvectors show large
values of $O_{ij}$. }
\label{overlap-daily}
\end{figure}

Next, we repeat our analysis for daily returns of 422 stocks using
$8685$ records of 1-day returns, and a sliding window of length
$L=965$ with discrete time steps $L/5=193$~days. Instead of calculating
{\bf \sf O($t,\tau$)} for all starting points $t$, we calculate {\bf
\sf O($\tau$)}$\equiv \langle$ {\bf \sf O($t, \tau$)} $\rangle_t$,
averaged over all $t=n\,L/5$, where $n\in \{0,1,2,\dots\}$.
Figure~\ref{overlap-daily} shows grey scale representations of {\bf
\sf O ($\tau$)} for increasing $\tau$. We find similar results as
found for shorter time scales, and find that eigenvectors
corresponding to the largest 2 eigenvalues are stable for time scales
as large as $\tau=$20~yr. In particular, the eigenvector {\bf \sf
u$^{422}$} shows an overlap of $\approx 0.8$ even over time scales of
$\tau=$30~yr.

\section{Applications to portfolio optimization}

\begin{figure}[hbt]
\narrowtext 
\centerline{
\epsfysize=0.9\columnwidth{\rotate[r]{\epsfbox{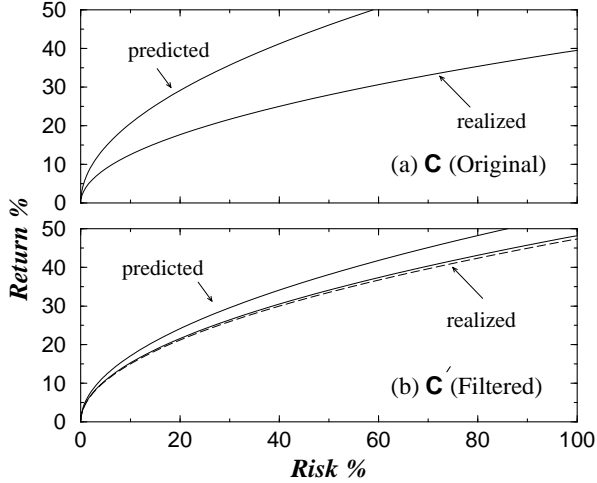}}}
}
\vspace{1cm}
\caption{(a) Portfolio return $R$ as a function of risk $D^2$ 
for the family of optimal portfolios (without a risk-free asset)
constructed from the original matrix {\bf \sf C}.  The top curve shows
the predicted risk $D_{\rm p}^2$ in 1995 of the family of optimal
portfolios for a given return, calculated using 30-min returns for
1995 and the correlation matrix {\bf
\sf C$_{94}$} for 1994.  For the same family of portfolios, the 
bottom curve shows the realized risk $D_{\rm r}^2$ calculated using
the correlation matrix {\bf \sf C$_{95}$} for 1995. These two curves
differ by a factor of $D_{\rm r}^2/D_{\rm p}^2\approx 2.7$.  (b)
Risk-return relationship for the optimal portfolios constructed using
the filtered correlation matrix {\bf \sf C$^{\prime}$}.  The top curve
shows the predicted risk $D_{\rm p}^2$ in 1995 for the family of
optimal portfolios for a given return, calculated using the filtered
correlation matrix {\bf \sf C$^{\prime}_{94}$}. The bottom curve shows
the realized risk $D_{\rm r}^2$ for the same family of portfolios
computed using {\bf \sf C$^{\prime}_{95}$}. The predicted risk is now
closer to the realized risk: $D_{\rm r}^2/D_{\rm p}^2\approx
1.25$. For the same family of optimal portfolios, the dashed curve
shows the realized risk computed using the original correlation matrix
{\bf \sf C$_{95}$} for which $D_{\rm r}^2/D_{\rm p}^2\approx 1.3$.}
\label{rr}
\end{figure}

The randomness of the ``bulk'' seen in the previous sections has
implications in optimal portfolio selection~\cite{Elton95}. We
illustrate these using the Markowitz theory of optimal portfolio
selection~\cite{Bouchaud00,aps-cizeau,Gopi00}. Consider a
portfolio $\Pi(t)$ of stocks with prices $S_i$. The return
on $\Pi(t)$ is given by
\begin{equation}
\Phi=\sum_{i=1}^{N} w_i G_i\,,
\label{retport}
\end{equation}
where $G_i(t)$ is the return on stock $i$ and $w_i$ is the fraction of
wealth invested in stock $i$. The fractions $w_i$ are normalized such
that $\sum_{i=1}^{N} w_i=1$.  The risk in holding the portfolio
$\Pi(t)$ can be quantified by the variance
\begin{equation}
\Omega^2=\sum_{i=1}^{N}\sum_{j=1}^N w_i w_j C_{ij} \sigma_i \sigma_j\,,
\label{varport}
\end{equation}
where $\sigma_i$ is the standard deviation (average volatility) of
$G_i$, and $C_{ij}$ are elements of the cross-correlation matrix {\bf
\sf C}.  In order to find an optimal portfolio, we must minimize
$\Omega^2$ under the constraint that the return on the portfolio is
some fixed value $\Phi$. In addition, we also have the constraint that
$\sum_{i=1}^N w_i =1$. Minimizing $\Omega^2$ subject to these two
constraints can be implemented by using two Lagrange multipliers,
which yields a system of linear equations for $w_i$, which can then be
solved.  The optimal portfolios thus chosen can be represented as a
plot of the return $\Phi$ as a function of risk $\Omega^2$
[Fig.~\ref{rr}].

To find the effect of randomness of {\bf \sf C} on the selected
optimal portfolio, we first partition the time period 1994--95 into
two one-year periods. Using the cross-correlation matrix {\bf
\sf C$_{94}$} for 1994, and $G_i$ for 1995, we construct a family of 
optimal portfolios, and plot $\Phi$ as a function of the predicted risk
$\Omega_{\rm p}^2$ for 1995 [Fig.~\ref{rr}(a)]. For this family of
portfolios, we also compute the risk $\Omega_{\rm r}^2$ realized during
1995 using {\bf \sf C$_{95}$} [Fig.~\ref{rr}(a)]. We find that the
predicted risk is significantly smaller when compared to the realized
risk, 
\begin{equation}
{\Omega_{\rm r}^2 - \Omega_{\rm p}^2 \over \Omega_{\rm p}^2}\approx 170\% \,.
\label{predict-risk}
\end{equation}

Since the meaningful information in {\bf \sf C} is contained in the
deviating eigenvectors (whose eigenvalues are outside the RMT bounds),
we must construct a `filtered' correlation matrix {\bf \sf
C$^\prime$}, by retaining only the deviating eigenvectors. To this
end, we first construct a diagonal matrix $\Lambda^\prime$, with
elements $\Lambda^{\prime}_{ii}=\{0,\dots,0,\lambda_{988},\dots,
\lambda_{1000}\}$. We then transform $\Lambda^\prime$ to the 
basis of {\bf \sf C}, thus obtaining the `filtered' cross-correlation
matrix {\bf \sf C$^\prime$}. In addition, we set the diagonal elements
$C^{\prime}_{ii}=1$, to preserve ${\rm Tr}$({\bf \sf C}) $=$ ${\rm
Tr}$({\bf \sf C}$^\prime$) $=N$. We repeat the above calculations for
finding the optimal portfolio using {\bf \sf C$^\prime$} instead of
{\bf \sf C} in Eq.~(\ref{varport}). Figure~\ref{rr}(b) shows that the
realized risk is now much closer to the predicted risk
\begin{equation}
{\Omega_{\rm r}^2 - \Omega_{\rm p}^2 \over \Omega_{\rm p}^2}\approx 25\% \,. 
\label{rmt-predict-risk}
\end{equation}
Thus, the optimal portfolios constructed using {\bf \sf C$^\prime$}
are significantly more stable in time.

\section{Conclusions}

How can we understand the deviating eigenvalues --- i.e., correlations
that are stable in time?  One approach is to postulate that returns
can be separated into idiosyncratic and common components --- i.e.,
that returns can be separated into different additive ``factors'',
which represent various economic influences that are common to a set
of stocks such as the type of industry, or the effect of
news~\cite{CLM,Sharpe70,Sharpe95,Sharpe64,Lintner65,Ross76,Brown85,Black72,Blume73,Fama92,FamaMacbeth73,RollRoss94,Chen86,Merton73a,Lehmann88,Campbell96,Campbell93a,Conner86a,Noh99,Marsili00,Cizeau00,Lillo00a}. 

On the other hand, in physical systems one starts from the
interactions between the constituents, and then relates interactions
to correlated ``modes'' of the system. In economic systems, we ask if
a similar mechanism can give rise to the correlated behavior. In order
to answer this question, we model stock price dynamics by a family of
stochastic differential equations~\cite{Farmer}, which describe the
`instantaneous'' returns $g_i(t)= {d\over dt}\ln S_i(t)$ as a random
walk with couplings $J_{ij}$
%
\begin{eqnarray}
\tau_o \partial_t g_i(t)= - r_i g_i(t) - \kappa g_i^3(t) + \sum_j J_{ij}
g_j(t) + {1\over \tau_o} \xi_i(t) \ \ \,.
\label{Langevin}
\end{eqnarray}
%
Here, $\xi_i(t)$ are Gaussian random variables with correlation
function $\langle \xi_i(t) \xi_j(t^\prime)\rangle = \delta_{ij} \tau_o
\delta(t-t^\prime)$, and $\tau_o$ sets the time scale of the problem.
In the context of a soft spin model, the first two terms in the {\it
rhs} of Eq.~(\ref{Langevin}) arise from the derivative of a
double-well potential, enforcing the soft spin constraint. The
interaction among soft-spins is given by the couplings $J_{ij}$. In
the absence of the cubic term, and without interactions, $\tau_o/r_i$
are relaxation times of the $\langle g_i(t) g_i(t+\tau) \rangle$
correlation function.
The return $G_i$ at a finite time interval $\Delta t$ is given by the
integral of $g_i$ over $\Delta t$.

Equation~(\ref{Langevin}) is similar to the linearized description of
interacting ``soft spins''~\cite{Fischer91} and is a generalized case
of the models of Refs.~\protect\cite{Farmer}.  Without interactions,
the variance of price changes on a scale $\Delta t \gg \tau_i$ is
given by $\langle (G_i(\Delta t))^2 \rangle= \Delta t / (r^2 \tau_i)$,
in agreement with recent studies~\cite{Plerou00}, where stock price
changes are described by an anomalous diffusion and the variance of
price changes is decomposed into a product of trading frequency
(analog of $1/\tau_i$) and the square of an ``impact parameter'' which
is related to liquidity (analog of $1/r$).

\begin{figure}[h]
\narrowtext 
\centerline{
\epsfysize=0.7\columnwidth{\rotate[r]{\epsfbox{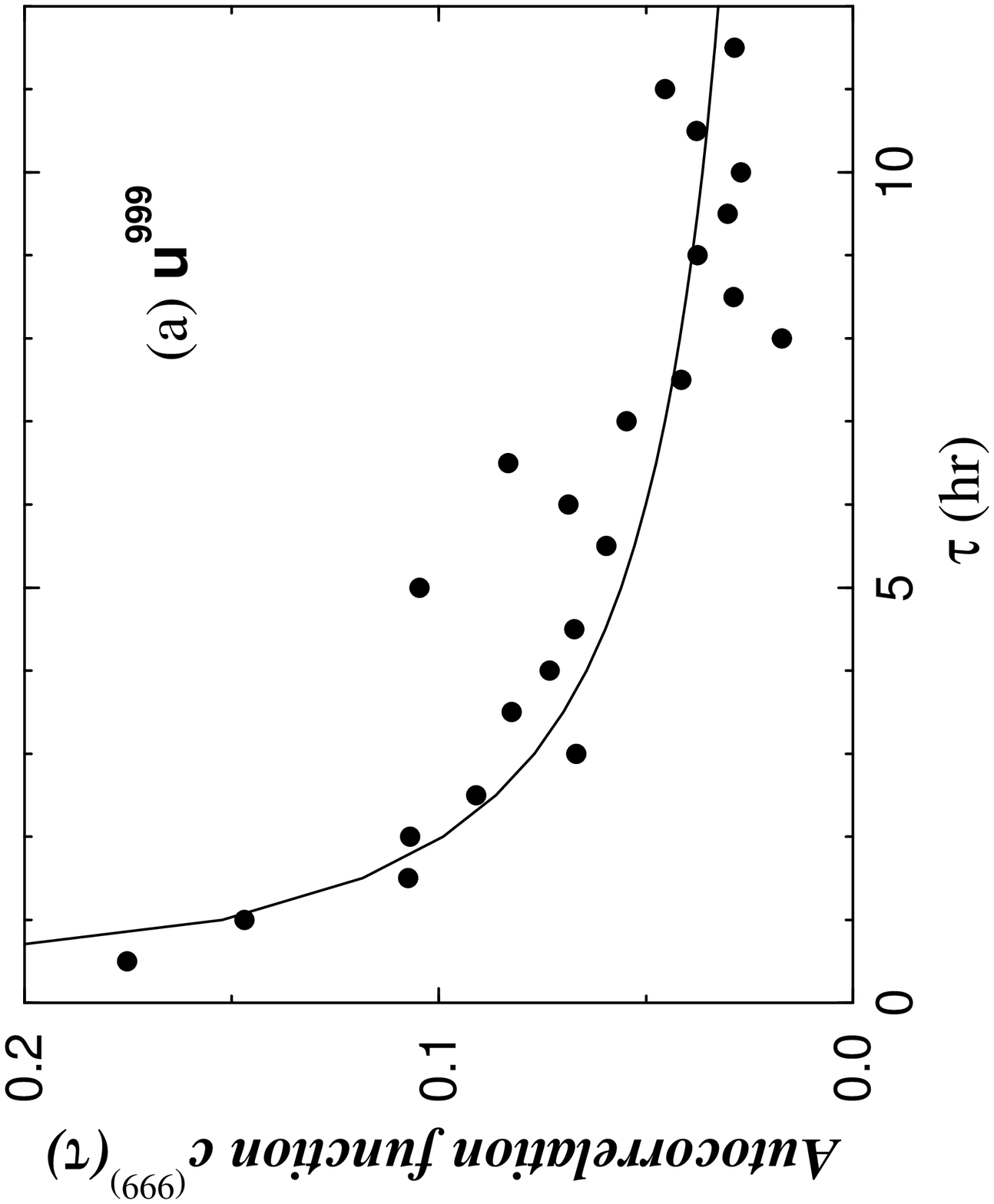}}} 
}
\vspace{0.2cm}
\centerline{
\epsfysize=0.7\columnwidth{\rotate[r]{\epsfbox{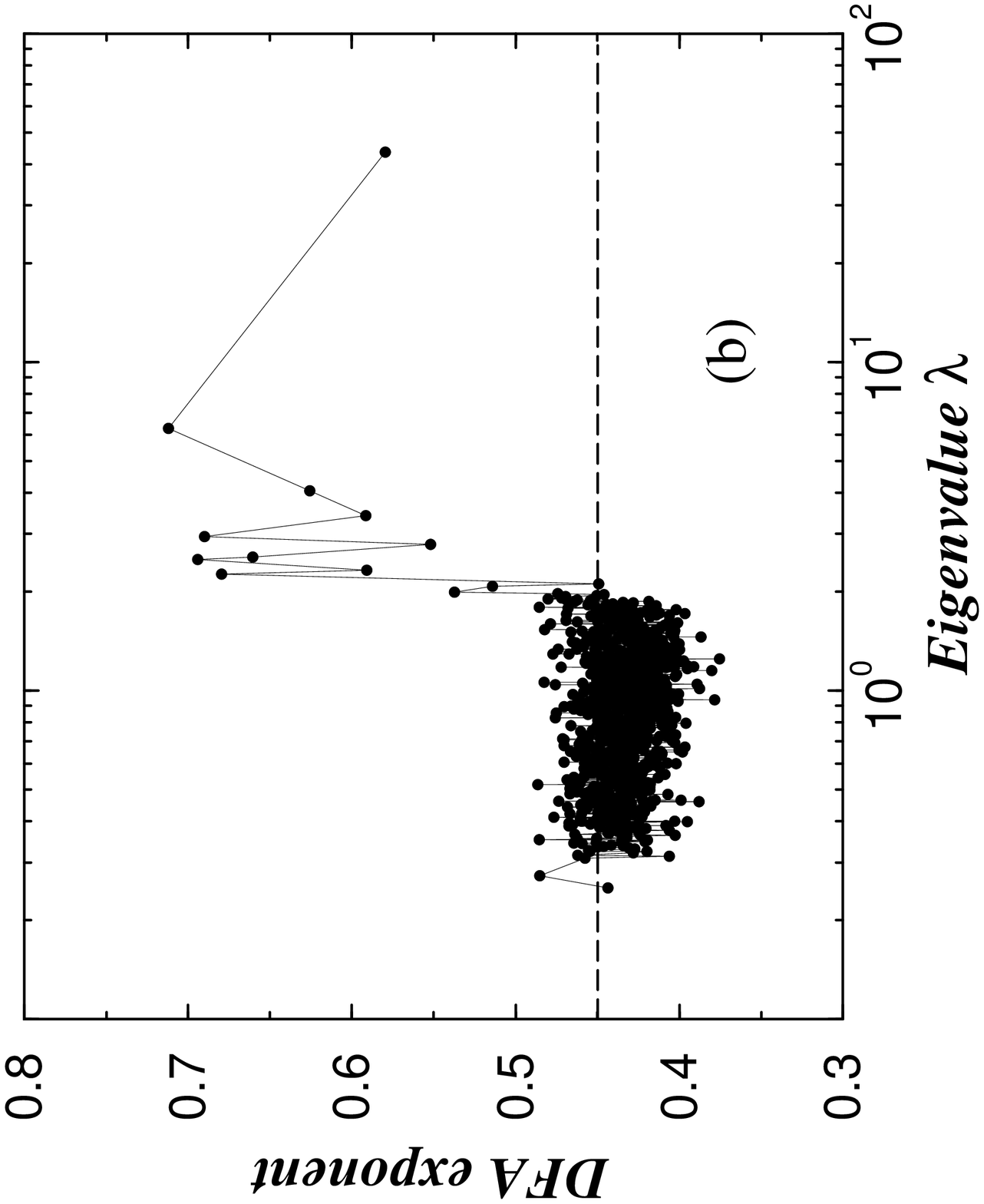}}} 
}
\vspace{0.5cm}
\caption{(a) Autocorrelation function $c^{(k)}(\tau)$ of the time series defined by
the eigenvector {\bf \sf u$^{999}$}. The solid line shows a fit to a
power-law functional form $\tau^{-\gamma_k}$, whereby we obtain values
$\gamma_k=0.61\pm0.06$. (b) To quantify the exponents $\gamma_k$ for all
$k=1,\dots,1000$ eigenvectors, we use the method of DFA
analysis~\protect\cite{Peng94} often used to obtain accurate estimates
of power-law correlations. We plot the detrended fluctuation function
$F(\tau)$ as a function of the time scale $\tau$ for each of the
$1000$ time series. Absence of long-range correlations would imply
$F(\tau) \sim \tau^{0.5}$, whereas $F(\tau) \sim \tau^{\nu}$ with $0.5
< \nu \leq 1$ implies power-law decay of the correlation function with
exponent $\gamma=2-2\nu$. We plot the exponents $\nu$ as a function of
the eigenvalue and find values exponents $\nu$ significantly larger
than $0.5$ for all the deviating eigenvectors. In contrast, for the
remainder of the eigenvectors, we obtain the mean value $\nu=0.44\pm
0.04$, comparable to the value $\nu =0.5$ for the uncorrelated case.}
\label{autocorr}
\end{figure}

As the coupling strengths increase, the soft-spin system undergoes a
transition to an ordered state with permanent local magnetizations. At
the transition point, the spin dynamics are very ``slow'' as reflected
in a power law decay of the spin autocorrelation function in time. To
test whether this signature of strong interactions is present for the
stock market problem, we analyze the correlation functions
$c^{(k)}(\tau)\equiv\langle G^{(k)}(t) G^{(k)}(t+\tau)\rangle$, where
$G^{(k)}(t)\equiv\sum_{i=1}^{1000} u_i^k G_i(t)$ is the time series
defined by eigenvector {\bf \sf u$^k$}.  Instead of analyzing
$c^{(k)}(\tau)$ directly, we apply the detrended fluctuation analysis
(DFA) method~\cite{Peng94}. Figure~\ref{autocorr} shows that the
correlation functions $c^{(k)}(\tau)$ indeed decay as power laws
~\cite{notesp} for the deviating eigenvectors {\bf \sf u$^k$} --- in
sharp contrast to the behavior of $c^{(k)}(\tau)$ for the rest of the
eigenvectors and the autocorrelation functions of individual stocks,
which show only short-ranged correlations.  We interpret this as
evidence for strong interactions~\cite{noteint}.
 
In the absence of the non-linearities (cubic term), we obtain only
exponentially-decaying correlation functions for the ``modes''
corresponding to the large eigenvalues, which is inconsistent with our
finding of power-law correlations.


To summarize, we have tested the eigenvalue statistics of the
empirically-measured correlation matrix {\bf \sf C} against the null
hypothesis of a random correlation matrix. This allows us to
distinguish genuine correlations from ``apparent'' correlations that
are present even for random matrices. We find that the bulk of the
eigenvalue spectrum of {\bf \sf C} shares universal properties with
the Gaussian orthogonal ensemble of random matrices. Further, we
analyze the deviations from RMT, and find that (i) the largest
eigenvalue and its corresponding eigenvector represent the influence
of the entire market on all stocks, and (ii) using the rest of the
deviating eigenvectors, we can partition the set of all stocks studied
into distinct subsets whose identity corresponds to
conventionally-identified business sectors. These sectors are stable
in time, in some cases for as many as 30~years.  Finally, we have seen
that the deviating eigenvectors are useful for the construction of
optimal portfolios which have a stable ratio of risk to return.

\section*{Acknowledgments}

We thank J-P. Bouchaud, S.~V. Buldyrev, P.~Cizeau, E.~Derman,
X.~Gabaix, J.~Hill, M.~Janjusevic, L.~Viciera, and J.~Zou for helpful
discussions. We thank O.~Bohigas for pointing out Ref.~\cite{Dyson71}
to us. BR thanks DFG grant RO1-1/2447 for financial support. TG thanks
Boston University for warm hospitality. The Center for Polymer Studies
is supported by the NSF, British Petroleum, the NIH, and the NRCPS
(PS1 RR13622).

\appendix
\section{``Unfolding'' the eigenvalue distribution}

As discussed in Section V, random matrices display {\it universal}
functional forms for eigenvalue correlations that depend only on the
general symmetries of the matrix. A first step to test the data for
such universal properties is to find a transformation called
``unfolding,'' which maps the eigenvalues $\lambda_i$ to new variables
called ``unfolded eigenvalues'' $\xi_i$, whose distribution is
uniform~\cite{Mehta91,Brody81,Guhr98}. Unfolding ensures that the
distances between eigenvalues are expressed in units of {\it local}
mean eigenvalue spacing~\cite{Mehta91}, and thus facilitates
comparison with analytical results.

We first define the cumulative distribution function of eigenvalues,
which counts the number of eigenvalues in the interval
$\lambda_i\le\lambda$,
\begin{equation}
F(\lambda) = N\,\int_{-\infty}^{\lambda} P (x) dx\,,
\label{staircase}
\end{equation}
where $P(x)$ denotes the probability density of eigenvalues and $N$ is
the total number of eigenvalues. The function $F(\lambda)$ can be
decomposed into an average and a fluctuating part,
\begin{equation}
F(\lambda)=F_{\rm av}(\lambda)+F_{\rm fluc}(\lambda)\,. 
\label{stair.dec}
\end{equation}
Since $P_{\rm fluc} \equiv dF_{\rm fluc}(\lambda)/d\lambda =0 $ on
average, 
\begin{equation}
P_{\rm rm}(\lambda)\equiv {dF_{\rm av}(\lambda) \over d\lambda}
\label{defrho}
\end{equation}
is the averaged eigenvalue density. The dimensionless, unfolded
eigenvalues are then given by
\begin{equation}
\xi_i \equiv F_{\rm av}(\lambda_i)\,.
\label{defunfold}
\end{equation}

Thus, the problem is to find $F_{\rm av}(\lambda)$. We follow two
procedures for obtaining the unfolded eigenvalues $\xi_i$: (i) a
phenomenological procedure referred to as Gaussian
broadening~\cite{Mehta91,Brody81,Guhr98}, and (ii) fitting the
cumulative distribution function $F(\lambda)$ of Eq.~(\ref{staircase})
with the analytical expression for $F(\lambda)$ using
Eq.~(\ref{densuncorr}). These procedures are discussed below.

\subsection{Gaussian Broadening}

Gaussian broadening~\cite{Brack72} is a phenomenological procedure
that aims at approximating the function $F_{\rm av}(\lambda)$ defined
in Eq.~\ref{stair.dec} using a series of Gaussian functions. Consider
the eigenvalue distribution $P(\lambda)$, which can be expressed as
\begin{equation}
P(\lambda) = {1 \over N} \sum_{i=1}^N \delta (\lambda -\lambda_i)\,.
\label{pdfdelta}
\end{equation}
The $\delta$-functions about each eigenvalue are approximated by
choosing a Gaussian distribution centered around each eigenvalue with
standard deviation $(\lambda_{k+a} - \lambda_{k-a})/2 $, where $2a$ is
the size of the window used for broadening~\cite{unfold}.  Integrating
Eq.~(\ref{pdfdelta}) provides an approximation to the function $F_{\rm
av}(\lambda)$ in the form of a series of error functions, which using
Eq.~(\ref{defunfold}) yields the unfolded eigenvalues.

\subsection{Fitting the eigenvalue distribution}

Phenomenological procedures are likely to contain artificial scales,
which can lead to an ``over-fitting'' of the smooth part $F_{\rm
av}(\lambda)$ by adding contributions from the fluctuating part
$F_{\rm fluc}(\lambda)$. The second procedure for unfolding aims at
circumventing this problem by fitting the cumulative distribution of
eigenvalues $F(\lambda)$ (Eq.~(\ref{staircase})) with the analytical
expression for 
\begin{equation}
F_{\rm rm}(\lambda)=N\,\int_{-\infty}^{\lambda} P_{\rm rm} (x) dx\,,
\label{sengupta-stair}
\end{equation}
where $P_{\rm rm}(\lambda)$ is the probability density of eigenvalues from
Eq.~(\ref{densuncorr}). The fit is performed with $\lambda_-$,
$\lambda_+$, and $N$ as free parameters. The fitted function is an
estimate for $F_{\rm av}(\lambda)$, whereby we obtain the unfolded
eigenvalues $\xi_i$. One difficulty with this method is that the
deviations of the spectrum of {\bf \sf C} from Eq.~(\ref{densuncorr})
can be quite pronounced in certain periods, and it is difficult to
find a good fit of the cumulative distribution of eigenvalues to
Eq.~(\ref{sengupta-stair}).


\begin{table}[hbt]
\narrowtext
\caption{Largest ten components of the eigenvectors {\bf \sf u$^{999}$} up
to {\bf \sf u$^{991}$}. The columns show ticker symbols, industry
type, and the Standard Industry Classification (SIC) code
respectively.}

\begin{tabular}{llc}
\multicolumn{1}{c}{Ticker} & \multicolumn{1}{c}{Industry} & \multicolumn{1}{c}{Industry Code} \\
\hline\\[2pt]
\multicolumn{3}{c}{{\bf \sf u$^{999}$}}\\
\hline
XON	&   Oil \& Gas Equipment/Services	&2911\\
PG	&   Cleaning Products		&2840\\
JNJ	&   Drug Manufacturers/Major 	&2834\\
KO	&   Beverages-Soft Drinks	&2080\\
PFE	&   Drug Manufacturers/Major	&2834\\
BEL	&   Telecom Services/Domestic	&4813\\
MOB	&   Oil \& Gas Equipment/Services	&2911\\
BEN	&   Asset Management		&6282\\
UN	&   Food - Major Diversified	&2000\\
AIG	&   Property/Casualty Insurance	&6331\\
\hline\\[2pt]
\multicolumn{3}{c}{{\bf \sf u$^{998}$}}\\
\hline
TXN	&   Semiconductor-Broad Line	&3674\\
MU	&   Semiconductor-Memory Chips	&3674\\
LSI	&   Semiconductor-Specialized	&3674\\
MOT	&   Electronic Equipment	&3663\\
CPQ	&   Personal Computers		&3571\\
CY	&   Semiconductor-Broad Line	&3674\\
TER	&   Semiconductor Equip/Materials	&3825\\
NSM	&   Semiconductor-Broad Line	&3674\\
HWP	&   Diversified Computer Systems	&3570\\
IBM	&   Diversified Computer Systems	&3570\\
\hline\\[2pt]
\multicolumn{3}{c}{{\bf \sf u$^{997}$}}\\
\hline
PDG	&   Gold			&1040\\
NEM	&   Gold			&1040\\
NGC	&   Gold			&1040\\
ABX	&   Gold			&1040\\
ASA	&   Closed-End Fund - (Gold)	&6799\\
HM	&   Gold			&1040\\
BMG	&   Gold			&1040\\
AU	&   Gold			&1040\\
HSM	&   General Building Materials	&5210\\
MU	&   Semiconductor-Memory Chips	&3674\\
\hline\\[2pt]
\multicolumn{3}{c}{{\bf \sf u$^{996}$}}\\
\hline
NEM	&   Gold			&1040\\
PDG	&   Gold			&1040\\
ABX	&   Gold			&1040\\
HM	&   Gold			&1040\\
NGC	&   Gold			&1040\\
ASA	&   Closed-End Fund - (Gold)	&6799\\
BMG	&   Gold			&1040\\
CHL	&   Wireless Communications	&4813\\
CMB	&   Money Center Banks		&6021\\
CCI	&   Money Center Banks		&6021\\
\hline\\[2pt]
\multicolumn{3}{c}{{\bf \sf u$^{995}$}}\\
\hline
TMX	&   Telecommunication Services/Foreign	&4813\\
TV	&   Broadcasting - Television		&4833\\
MXF	&   Closed-End Fund - Foreign	&6726\\
ICA	&   Heavy Construction		&1600\\
GTR	&   Heavy Construction		&1600\\
CTC	&   Telecom Services/Foreign	&4813\\
PB	&   Beverages-Soft Drinks	&2086\\
YPF	&   Independent Oil \& Gas	&2911\\
TXN	&   Semiconductor-Broad Line	&3674\\
MU	&   Semiconductor-Memory Chips	&3674\\
\hline\\[2pt]
\multicolumn{3}{c}{{\bf \sf u$^{994}$}}\\
\hline
BAC	&   Money Center Banks		&6021\\
CHL	&   Wireless Communications	&4813\\
BK	&   Money Center Banks		&6022\\
CCI	&   Money Center Banks		&6021\\
CMB	&   Money Center Banks		&6021\\
BT	&   Money Center Banks		&6022\\
JPM	&   Money Center Banks		&6022\\
MEL	&   Regional-Northeast Banks	&6021\\
NB	&   Money Center Banks		&6021\\
WFC	&   Money Center Banks		&6021\\
\hline\\[2pt]
\multicolumn{3}{c}{{\bf \sf u$^{993}$}}\\
\hline
BP	&   Oil \& Gas Equipment/Services	&2911\\
MOB	&   Oil \& Gas Equipment/Services	&2911\\
SLB	&   Oil \& Gas Equipment/Services	&1389\\
TX	&   Major Integrated Oil/Gas	&2911\\
UCL	&   Oil \& Gas Refining/Marketing	&1311\\
ARC	&   Oil \& Gas Equipment/Services	&2911\\
BHI	&   Oil \& Gas Equipment/Services	&3533\\
CHV	&   Major Integrated Oil/Gas	&2911\\
APC	&   Independent Oil \& Gas	&1311\\
AN	&   Auto Dealerships		&2911\\
\hline\\[2pt]
\multicolumn{3}{c}{{\bf \sf u$^{992}$}}\\
\hline
FPR	&   Auto Manufacturers/Major	&3711\\
F	&   Auto Manufacturers/Major	&3711\\
C	&   Auto Manufacturers/Major	&3711\\
GM	&   Auto Manufacturers/Major	&3711\\
TXN	&   Semiconductor-Broad Line	&3674\\
ADI	&   Semiconductor-Broad Line	&3674\\
CY	&   Semiconductor-Broad Line	&3674\\
TER	&   Semiconductor Equip/Materials	&3825\\
MGA	&   Auto Parts			&3714\\
LSI	&   Semiconductor-Specialized	&3674\\
\hline\\[2pt]
\multicolumn{3}{c}{{\bf \sf u$^{991}$}}\\
\hline
ABT&   Drug Manufacturers/Major		&2834\\
PFE&   Drug Manufacturers/Major		&2834\\
SGP&   Drug Manufacturers/Major		&2834\\
LLY&   Drug Manufacturers/Major		&2834\\
JNJ&   Drug Manufacturers/Major		&2834\\
AHC&   Oil \& Gas Refining/Marketing	&2911\\
BMY&   Drug Manufacturers/Major		&2834\\
HAL&   Oil \& Gas Equipment/Services	&1600\\
WLA&   Drug Manufacturers/Major		&2834\\
BHI&   Oil \& Gas Equipment/Services	&3533\\
\end{tabular}
\label{industry.tab}
\end{table}

\end{multicols}


\begin{thebibliography}{99}
\bibitem{splits}
The time series of prices have been adjusted for stock splits and
dividends.
\bibitem{noteuniv}
Analytical evidence for this ``universality'' is summarized in
Section~8 of Ref.~\protect\cite{Guhr98}.
\bibitem{notenn}
For GOE matrices, the Wigner surmise (Eq.~(\ref{eq.3})) is not
exact~\protect\cite{Mehta91}. Despite this fact, the Wigner surmise is
widely used because the difference between the exact form of $P_{\rm
GOE}(s)$ and Eq.~(\ref{eq.3}) is almost
negligible~\protect\cite{Mehta91}.
\bibitem{notegse}
Analogous to the GOE, which is defined on the space of real symmetric
matrices, one can define two other ensembles~\cite{Mehta91}: (i) the
Gaussian unitary ensemble (GUE), which is defined on the space of
hermitian matrices, with the requirement that the joint probability of
elements is invariant under unitary transformations, and (ii) the
Gaussian symplectic ensemble (GSE), which is defined on the space of
hermitian ``self-dual'' matrices with the requirement that the joint
probability of elements is invariant under symplectic transformations.
Formal definitions can be found in Ref.~\protect\cite{Mehta91}.
\bibitem{noteint}
For the case of predominantly ``ferromagnetic'' couplings ($J_{ij}
>0$) within disjoint groups of stocks, a factor model (such as
Eq.~(\protect\ref{marketm})) can be derived from the model of
``interacting stocks'' in Eq.~(\ref{Langevin}). In the spirit of a
mean-field approximation, the influence of the price changes of all
other stocks in a group on the price of a given stock can be modeled
by an effective field, which has to be calculated
self-consistently. This effective field would then play a similar role
as a factor in standard economic models.
\end{thebibliography}
\end{document}